\documentclass[twoside]{article}
\usepackage{graphicx} % Required for inserting images
\usepackage{tikz-cd}
\usepackage{hyperref}
\hypersetup{colorlinks,
citecolor=blue
}

\usepackage{float}

\usepackage{amsmath,amsfonts,amscd,amssymb,array, amsthm,mathtools}
\usepackage{authblk}
\usepackage{blindtext}
\usepackage{xcolor}

\usepackage{fancyhdr}

\newenvironment{acknowledgements}{%
  \begin{abstract}
}{%
  \end{abstract}
}

\usepackage [n, lambda, advantage, operators, sets,
adversary, landau, probability, notions,
logic, ff, mm, primitives, events, complexity, oracles, asymptotics, keys]{cryptocode}

\usepackage{algorithm}
\usepackage{algpseudocode}
\usepackage{multicol}

\usepackage{tabularx}

\newcounter{protocol}

\newtheorem{theorem}{Theorem}[section]
\newtheorem{lemma}[theorem]{Lemma}
\newtheorem{remark}[theorem]{Remark}

\newtheorem{problem}[theorem]{Problem}

\newtheorem{definition}[theorem]{Definition}
\newtheorem{claim}[theorem]{Claim}

\DeclareMathOperator{\groupendo}{End}

\title{\textsf{SQIAsignHD}: SQIsignHD Adaptor Signature}
\author[1]{Farzin Renan}
\author[2]{Péter Kutas}
\affil[1]{Middle East Technical University, Ankara, Turkey}
\affil[ ]{$\mathsf{ farzin.renan@gmail.com}$}
\affil[ ]{ }
\affil[2]{University of Birmingham, UK}
\affil[2]{Eötvös Loránd University, Budapest, Hungary}
\affil[ ]{$\mathsf{p.kutas@bham.ac.uk}$}

\date{}

\begin{document}
% Set the page style to "fancy"...

%... then configure it.
\fancyhead{} % clear all header fields
\fancyfoot{} % clear all header fields
\fancyhead[OR]{\textsf{SQIAsignHD}: SQIsignHD Adaptor Signature \hspace{.5cm} \thepage}
\fancyhead[EL]{\thepage \hspace{.5cm} Farzin Renan and P\'eter Kutas}

\maketitle

%\tableofcontents
%\newpage
\begin{abstract}
  Adaptor signatures can be viewed as a generalized form of standard digital signature schemes by linking message authentication to the disclosure of a secret value. As a recent cryptographic primitive, they have become essential for blockchain applications, including cryptocurrencies, by reducing on-chain costs, improving fungibility, and enabling off-chain payments in payment-channel networks, payment-channel hubs, and atomic swaps. However, existing adaptor signature constructions are vulnerable to quantum attacks due to Shor’s algorithm. 
   In this work, we introduce $\mathsf{SQIAsignHD}$, a new quantum-resistant adaptor signature scheme based on isogenies of supersingular elliptic curves, using SQIsignHD - as the underlying signature scheme - and exploiting the idea of the artificial orientation on the supersingular isogeny Diffie-Hellman key exchange protocol, SIDH, to define the underlying hard relation. We, furthermore, provide a formal security proof for our proposed scheme.\\
   
   \noindent\begin{keywords}
       {Post-quantum Cryptography, Blockchain, Isogeny-based Cryptography, Adaptor Signature, Payment Channel Network}
   \end{keywords}
\end{abstract}
\begin{acknowledgements}
      \noindent We express our gratitude to Luca De Feo, Andrea Basso, Simon-Philipp Merz, and the cryptography research group at IBM for their valuable feedback. We also thank the anonymous reviewers for their insightful comments, which helped improve this paper.
   \end{acknowledgements}
   
\section{Introduction}\label{sec1}

Blockchain technology, introduced anonymously in 2009 \cite{bitcoin}, revolutionized digital payments by enabling decentralized financial transactions recorded in a distributed ledger. Each transaction is validated by network nodes through a consensus protocol, forming the backbone of cryptocurrencies such as Bitcoin and Ethereum. However, executing transactions on-chain incurs fees based on storage and computational costs, making frequent transactions expensive. To address this, off-chain solutions were explored to reduce on-chain fees while preserving security. In this context, Andrew Poelstra introduced the concept of scriptless scripts \cite{poelstra}, which was later formalized as adaptor signatures by \cite{aumayr} and \cite{formalizedAdaptor}, providing a more efficient mechanism for conditional payments without relying on complex on-chain scripts.

\subsection{Adaptor Signature}

An adaptor signature is a novel cryptographic primitive that builds upon the concept of a standard digital signature. It has emerged as a key tool for blockchain applications, such as cryptocurrencies, to reduce on-chain costs, improve fungibility, and support off-chain payment methods in payment-channel networks (PCNs), payment-channel hubs (PCHs), and atomic swaps. Adaptor signatures also play a crucial role in Anonymous Multihop Locks (AMHLs), enabling secure and private conditional transfers by embedding a secret within the signature. This feature ensures that transactions in AMHLs remain atomic and conditional on the revelation of the secret \cite{malavolta}.

Technically, an adaptor signature conceals secret randomness by embedding it within the signature during the signing process. This randomness is revealed once the signature is created. Specifically, the typical procedure involves constructing a pre-signature in the first phase, converting it into a full signature using secret randomness in the second phase, and finally extracting the secret randomness from the signature using cryptographic processing. Furthermore, the signature produced by an adaptor signature can be verified using the verification algorithm of the underlying signature scheme.

An adaptor signature also possesses features that ensure its security. A signer with a secret key can create a pre-signature for any message, which can then be converted into a full signature if and only if the user possesses a valid witness to the statement. Furthermore, anyone with access to both the pre-signature and the corresponding full signature can extract the witness and reveal the hard relation.

\subsection{Related Work and Our Contribution}

Several works have explored adaptor signatures and their applications. Aumayr et al. \cite{aumayr} provide a formalization of adaptor signatures, applying them to ECDSA and Schnorr-based schemes. Malavolta et al. \cite{malavolta} analyze secure and privacy-preserving PCNs, identifying a new attack that affects major PCNs, such as the Lightning Network. They also define Anonymous Multihop Locks (AMHLs) and demonstrate how they can be constructed for PCNs using linear homomorphic one-way functions. Moreno-Sanchez et al. \cite{moreno} show an instance of adaptor signatures applied to Monero’s linkable ring signature scheme to improve scalability and address other issues. Tairi et al. \cite{tairi} introduce the PCHs protocol, with a provably secure instantiation based on adaptor signatures. However, these constructions are vulnerable to quantum adversaries due to Shor’s algorithm \cite{shor}.

The security of blockchain technologies largely depends on digital signature schemes built on Elliptic Curve Cryptography (ECC) to authenticate transactions. ECC's security relies on the intractability of the discrete logarithm problem, which is secure against classical computers. However, Shor’s algorithm enables quantum computers to efficiently compute discrete logarithms in polynomial time. Additionally, due to Grover’s algorithm \cite{grover}, quantum attackers could potentially replace valid blocks with falsified ones, making blockchains susceptible to quantum attacks. In the case of Bitcoin, for instance, this could allow attackers to double-spend or steal assets from other users. As a result, post-quantum cryptography has gained increasing attention and has become a critical area of research. To secure cryptosystems against quantum adversaries, the underlying hard problems must remain intractable in the quantum setting.

In the realm of post-quantum cryptography, the first post-quantum adaptor signature, LAS \cite{LAS}, was established using lattice-based assumptions such as Module-LWE and Module-SIS, with a simplified form of Dilithium \cite{dilithium} as the underlying signature scheme. Applications using LAS require zero-knowledge proofs to ensure the extracted witness satisfies the desired norm and the hard relation. However, the most efficient proof variant is 53KB \cite{53K}, leading to significant off-chain communication costs. Moreover, LAS, when used in specific applications like PCNs, can leak non-trivial information, compromising the overall privacy of the architecture.

Another attempt at designing an adaptor signature, named SQI-AS, was introduced in \cite{val}, using SQISign \cite{sqisign} as the underlying signature. The authors rely on SIDH \cite{sidh} to apply the corresponding hard relation in their design. However, due to devastating attacks \cite{CDattackSIDH, attack2SIDH, attack3} on SIDH, SQI-AS lost its security. This vulnerability arises because SQI-AS’s adapting algorithm benefits from SIDH-like operations, requiring the publication of torsion point images as auxiliary information during the pre-signature phase. This SIDH-based information is critical for breaking SIDH security and exposing the secret key isogeny. Furthermore, SQI-AS suffers from structural flaws in its design and security proof. Specifically, the shifting of the signature with secret randomness does not occur correctly. In an adaptor signature, any two of the trio (witness, pre-signature, and signature) must generate the others; however, in SQI-AS, there is no mechanism to generate the pre-signature from the witness and signature. As a result, generating the pre-signature from the full signature and witness, which is necessary for the simulator $\mathcal{S}$ of $\mathsf{NIZK}$ to simulate oracle queries using the signing oracle $\mathsf{Sig^{SQISign}}$ and the random oracle $\mathcal{H}^\mathsf{SQISign}$ for adversary $\mathcal{A}$, becomes inapplicable. Furthermore, in SQI-AS, the generated signature is not directly verifiable using the standard verification procedure of the underlying signature scheme.

The only secure isogeny-based adaptor signature scheme in the literature is IAS \cite{ias}, which uses CSI-FiSh \cite{csifish} as the underlying signature and relies on the security of the CSIDH key exchange protocol \cite{csidh}. However, IAS’s efficiency is limited by the parameter sizes of CSI-FiSh. Specifically, CSI-FiSh operates with a maximum of CSIDH-512 parameters since knowledge of the class group structure is required to efficiently compute the class group action on random group elements. CSIDH-512 is relatively slow and vulnerable to quantum subexponential attacks. Recent quantum algorithms \cite{bonnetain, peikert} have demonstrated that the parameters of CSIDH-512 do not provide the required quantum security, leading to ongoing debates about their adequacy. A new isogeny-based group action, named SCALLOP and proposed by De Feo et al. \cite{scallop}, addresses the scaling problem with CSI-FiSh. SCALLOP simplifies the computation of the class group structure but requires more computations to execute the group action, making it slower than CSI-FiSh. 

\paragraph{Contribution.} In light of these challenges, this work introduces a new post-quantum adaptor signature based on SQIsignHD \cite{sqisignhd}, the most compact post-quantum digital signature available. Compared to other isogeny-based signature schemes, SQIsignHD is generally faster and more flexible in its parameter sets. Therefore, unlike IAS, which is restricted to CSIDH-512 parameters and is susceptible to quantum subexponential attacks, our scheme scales well to higher security levels. The signature in our construction is approximately $1.26$ KB in size for a $\lambda=128$ security level. 

The main technical challenges in constructing isogeny-based adaptor signatures stem from the fact that not all post-quantum digital signatures, particularly SQIsignHD, satisfy certain homomorphic properties. As shown by \cite{adaptor1}, signature schemes derived from identification (ID) schemes with homomorphic features can be generically transformed into adaptor signature schemes. To address this, we carefully apply the concept of “shifting the signature by secret randomness” using several techniques, allowing SQIsignHD to meet this requirement. We also leverage recent advances in SIDH attacks to recover the secret witness during the extraction phase of our construction.

\subsection{Organization of the Paper}

Section \ref{sec2} provides the necessary preliminaries for the main sections, Sections \ref{sec3} and \ref{sec4}. These preliminaries are divided into two parts: the mathematical prerequisites for our construction and the cryptographic background required for the next sections. Section \ref{sec3} introduces the new adaptor signature $\mathsf{SQIAsignHD}$ and examines it in detail. Section \ref{sec4} analyzes the security of $\mathsf{SQIAsignHD}$, providing a formal proof of its security in the random oracle model.

\section{Preliminaries}\label{sec2}

\noindent\textbf{Notation.} A \textit{negligible} function $\mathsf{negl}: \mathbb{N} \rightarrow \mathbb{R}$ is a function that, for every $k \in \mathbb{N}$, admits $\mathcal{O}(n^{-k})$ as its upper bound, i.e., there exists $n_0 \in \mathbb{N}$ such that for every $n \geq n_0$, it holds that $\mathsf{negl}(n) \leq 1/n^k$. We denote the uniform sampling of the variable $x$ from the set $X$ by $x \xleftarrow{\$} X$. Moreover, we denote a probabilistic polynomial-time ($\mathsf{PPT}$) algorithm $A$ on input $y$, producing output $x$, by $x \xleftarrow{\$} A(y)$. If the algorithm $A$ is deterministic polynomial-time ($\mathsf{DPT}$), it is denoted by $x := A(y)$.

\subsection{Elliptic Curves and Isogenies} 

\textbf{Elliptic Curves.} Let $k := \mathbb{F}_q$ be a finite field where $q = p^n$ for some prime $p$ and positive integer $n$, with $\operatorname{char}(k) = p \neq 2, 3$. An \emph{elliptic curve} $E$, over a field $k$, is a smooth projective curve of genus 1, defined over $k$, with a distinguished $k$-rational point $\infty:= [0:1:0]$. Every elliptic curve over field $k$ can be uniquely represented (up to $\bar{k}$-isomorphism) by its $j$-invariant. For a positive integer $l$, the \emph{$l$-torsion subgroup} of $E$ is defined as $E[l] := \{ P \in E(\overline{k}) \mid [l]P = \infty \}$. An elliptic curve $E$ is said to be \emph{supersingular} if it has no nontrivial $p$-torsion points over  $\overline{\mathbb{F}}_p$, i.e., $E[p] = \{\infty\}$. If $E$ is supersingular, then $\operatorname{char}(k) = p$ divides  $|E(\mathbb{F}_q)| - q - 1$.\\

\noindent\textbf{Isogenies.} An \textit{isogeny} $\varphi: E_1 \to E_2 $ is a surjective morphism that maps the point at infinity of $E_1$ to the point at infinity of $E_2$. Two elliptic curves $E_1$ and $E_2$ are \textit{isogenous} over $\mathbb{F}_q$ if there exists an isogeny between them over $\mathbb{F}_q$. Furthermore, Tate’s theorem \cite{silverman} says that $E_1$ and $E_2$ are isogenous over $\mathbb{F}_q$ if and only if $|E_1(\mathbb{F}_q)| = |E_2(\mathbb{F}_q)|$. The \textit{degree} of isogeny $\varphi$ is the degree of 
the field extension $[k(E_1) : \varphi^{\ast} (k(E_2))]$, where $k(E_i)$ is the function field of $E_i$, $i=1,2$, and $\varphi^{\ast}$ is the pullback of $\varphi$ defined as $\varphi^\ast: k(E_2)\rightarrow k(E_1)$, with $\varphi^\ast(f) := f \circ \varphi$ for $f \in k(E_2)$. The isogeny $\varphi$ is called \textit{separable} in case the field extension is separable. If $\gcd(\deg(\varphi), \text{char}(k))= 1$, then $\varphi$ is necessarily separable. Since $\varphi(\infty_{E_1}) = \infty_{E_2}$, it follows that $\varphi:E_1(k) \to E_2(k)$ is a group homomorphism. If $\varphi$ is separable, then $|\ker(\varphi)| = \deg(\varphi)$.  Therefore, isogenies can be characterized by their kernel. In particular, there is a one-to-one correspondence between separable isogenies (up to isomorphism of the target curve) and finite subgroups of $E_1 (k)$. Isogenies can be constructed from their kernels using Vélu’s formulas \cite{velu}. Such an isogeny takes the form $E\rightarrow E/G$, where $G$ is a finite subgroup of $E$, and the kernel of the constructed isogeny. Since the degree of a composition of isogenies equals the product of their degrees, for any isogeny $\phi$ of degree $l= \prod_{i=1}^n l_i $, $\phi$ can be factored as a composition of $l_i$-isogenies, where $1\leq i\leq n$ and the integers $l_i$ need not be coprime. 

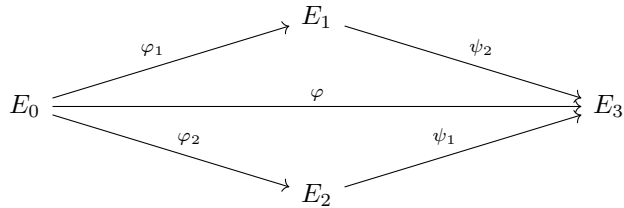
\begin{figure}[h]
\centering
\begin{tikzcd}[column sep=90pt]
    & E_1\arrow[dr, "\psi_2"] \\
    E_0 \arrow[ur, "\varphi_1"] \arrow[dr, "\varphi_2"] \arrow[rr, "\varphi"]& & E_3 \\
    & E_2 \arrow[ur, "\psi_1"]
\end{tikzcd}
\caption{Commutative Isogeny Diagram.}
\label{fig:commdiag}
\end{figure}

If the $l_i$'s are pairwise coprime, then reordering the $l_i$'s produces a different set of isogenies due to the non-commutative structure of isogenies of supersingular elliptic curves under composition. Suppose that $l_1$ and $l_2$ are two coprime integers and $\varphi$ is an $l_{1}l_{2}$-isogeny. Then, $\varphi$ can be decomposed in two ways: $\varphi= \psi_2\circ \varphi_1=\psi_1\circ\varphi_2$, as shown in Figure \ref{fig:commdiag}. In this case, $\psi_1$ (respectively $\psi_2$) is called the \textit{push-forward} of $\varphi_1$ (respectively $\varphi_2$) through $\varphi_2$ (respectively $\varphi_1$), denoted by $\psi_1 = [\varphi_2]_\ast\varphi_1$ (respectively $\psi_2 = [\varphi_1]_\ast \varphi_2$). It can be shown that ker$(\psi_1)=\varphi_2 (\text{ker}(\varphi_1))$, and ker$(\psi_2)=\varphi_1 (\text{ker}(\varphi_2))$. Furthermore, $\varphi_1$ (respectively $\varphi_2$) is called the \textit{pull-back} of $\psi_1$ (respectively $\psi_2$) through $\varphi_2$ (respectively $\varphi_1$), denoted by $\varphi_1 = [\varphi_2]^\ast\psi_1$ (respectively $\varphi_2 = [\varphi_1]^\ast \psi_2$).

For a given isogeny $\alpha : E_1 \to E_2$ of degree $d$, its (unique) \textit{dual} is an isogeny $\hat{\alpha}: E_2 \to E_1$ of degree $d$ such that $\alpha \circ \hat{\alpha} = [d]:E_2 \rightarrow E_2$, and $\hat{\alpha} \circ\alpha = [d]:E_1 \rightarrow E_1$. An isogeny from an elliptic curve $E$ to itself is called an \textit{endomorphism}. Notable examples of endomorphisms include the \textit{multiplication-by-integer-$m$} map $[m]:P\mapsto m\cdot P$, and the \textit{Frobenius} map $\pi : (x, y) \mapsto (x^q, y^q )$ of an elliptic curve defined over $E/\mathbb{F}_q$. The set of all endomorphisms on $E$, denoted by $\groupendo(E)$, forms a ring under addition and composition, known as the \textit{endomorphism ring} of $E$. Every supersingular elliptic curve in characteristic $p$ is isomorphic to a supersingular elliptic curve defined over $\mathbb{F}_{p^2}$. This implies that each supersingular elliptic curve has an isomorphic representative defined over $\mathbb{F}_{p^2}$. For a prime $\ell\neq p$, the supersingular $\ell$-isogeny graph is the graph whose vertices represent the supersingular $j$-invariants in $\mathbb{F}_{p^2}$, and whose edges correspond to the $\ell$-isogenies between them. These graphs are connected \cite{Kohel}, essentially undirected (since each $\ell$-isogeny has a dual), $(\ell+1)$-regular (since there are exactly $\ell+1$ outgoing edges from each $j$-invariant), and Ramanujan \cite{ramanujan}.

\subsection{Endomorphism Rings and Quaternion Orders} 

\textbf{Quaternion Algebras.} Let $a, b \in \mathbb{Q}^*$. A \textit{quaternion algebra} $\mathcal{B}$ over $\mathbb{Q}$ is a four-dimensional central simple $\mathbb{Q}$-algebra defined as $\mathcal{B}:= (\frac{a,b}{\mathbb{Q}})=\mathbb{Q} +\mathbb{Q}i+\mathbb{Q}j+\mathbb{Q}k$, where $1, i, j, k$ form a basis satisfying $i^2 = a$, $j^2 = b$, and $k = ij = -ji$. Let $l$ be a prime. The quaternion algebra $\mathcal{B}_l:= \mathcal{B} \otimes_\mathbb{Q} \mathbb{Q}_l$ is obtained by extending the scalars of $\mathcal{B}$ from $\mathbb{Q}$ to $\mathbb{Q}_l$, where $\mathbb{Q}_l$ is the set of $l$-adic numbers (i.e., the fraction field of $l$-adic integers $\mathbb{Z}_l$ which is the localization of $\mathbb{Z}$ away from prime $l$). Also, we can define $\mathcal{B}_\infty:= \mathcal{B} \otimes_\mathbb{Q} \mathbb{R}$. A quaternion algebra $\mathcal{B}$ is said to be ramified at $l$ (including $l=\infty$) if $\mathcal{B}_l$ is a division algebra. We are only interested in $\mathcal{B}_{p,\infty}$ which is a quaternion algebra ramified at $p$ and $\infty$. A \textit{fractional ideal} $I$ of $\mathcal{B}$ is a $\mathbb{Z}$-lattice of rank four, expressible as $I = \mathbb{Z}\alpha_1 + \mathbb{Z}\alpha_2+ \mathbb{Z}\alpha_3 + \mathbb{Z}\alpha_4$, for some $\mathbb{Q}$-basis $\{ \alpha_1, \alpha_2, \alpha_3, \alpha_4\}$ of $\mathcal{B}$.\\

\noindent\textbf{Quaternionic Orders.} An \textit{order} is a fractional ideal that is also a subring of $\mathcal{B}$. An order $\mathcal{O}$ is \textit{maximal}  if it is not strictly contained in any other order. Let $E$ be an elliptic curve defined over a field of characteristic $p$ with no non-trivial $p$-torsion points, namely supersingular. The endomorphism algebra of such an elliptic curve is isomorphic to a quaternion algebra ramified at $p$ and $\infty$, and its endomorphism ring is isomorphic to a maximal order of the corresponding quaternion algebra, i.e., $\groupendo^0(E):= \groupendo(E)\otimes_\mathbb{Z} \mathbb{Q}\cong \mathcal{B}_{p,\infty}$, and $\groupendo(E)\cong \mathcal{O}\subset \mathcal{B}_{p,\infty}.$
Conversely, for any maximal order in $\mathcal{B}_{p,\infty}$, there exists a supersingular elliptic curve over a field of characteristic $p$ such that whose endomorphism ring is isomorphic to this maximal order. This correspondence, known as the \textit{Deuring correspondence} \cite{deuring}, establishes a connection between supersingular elliptic curves and maximal orders in quaternion algebras. Specifically, given a fixed maximal order \(\mathcal{O} \cong \operatorname{End}(E)\), there exists an equivalence between the category of supersingular elliptic curves (under isogenies) and the category of left fractional $\mathcal{O}$-ideals (under homomorphisms of $\mathcal{O}$-modules). Constructing a supersingular elliptic curve with a given maximal order as its endomorphism ring (one direction of the Deuring correspondence) is computationally feasible in polynomial time over carefully chosen base fields \cite{klpt}. This procedure is known as the \textit{constructive Deuring correspondence} \cite{people}. Let $\mathcal{O} \subset \mathcal{B}_{p,\infty}\cong\groupendo^{0}(E)$ be a maximal order, and let $I$ be an integral left $\mathcal{O}$-ideal. The set of $I$-torsion points of $E$ is defined as $E[I]:= \{P \in E: \alpha(P)=0, \text{ for all } \alpha \in I\}$, which corresponds to the kernel of $I$. For such an ideal $I$, the associated isogeny $\varphi_I: E \to E_I:=\Large{\frac{E}{E[I]}}$ is defined with kernel $E[I]$.

\subsection{Artificial Orientation}

Artificial orientation, introduced in \cite{artificial}, provides a method for securely computing SIDH-like operations to counteract current SIDH attacks. Let $A$ and $B$ be smooth, square-free, and relatively prime integers, and let $p$ be a prime of the form $p = ABf - 1$, where $f$ is a small cofactor. Let $E$ be a supersingular elliptic curve defined over $\mathbb{F}_{p^2}$. An \textit{artificial $\mathit{A}$-orientation} of $E$ is a pair $\mathfrak{A} = (G_1, G_2)$, where $G_1$, $G_2$ are cyclic subgroups of $E[A]$ satisfying $|G_1| = |G_2| = A$ and $G_1 \cap G_2 = \{\infty\}$. A curve $E$ equipped with $\mathfrak{A}$ is called an \textit{artificially $\mathit{A}$-oriented curve}, denoted $(E, \mathfrak{A})$. For an artificially $A$-oriented curve $(E, \mathfrak{A})$, a range of isogenies can be constructed with kernels derived from $\mathfrak{A} = (G_1, G_2)$. Specifically, an isogeny $\phi$ is termed an $\mathfrak{A}\textit{-isogeny}$ if its kernel can be expressed as $\ker(\phi) = H_1 \oplus H_2$, where $H_1 \subseteq G_1$ and $H_2 \subseteq G_2$. Such an isogeny can be decomposed into two isogenies of relatively prime degrees as $\phi = \phi_2 \circ \phi_1$, where $\ker(\phi_1) = H_1 \subseteq G_1$ and $\ker(\phi_2) = \phi_1(H_2) \subseteq \phi_1(G_2)$.

However, as noted in \cite{artificial}, for a non-trivial $\mathfrak{A}$-isogeny $\phi: E \to E'$, the artificial $A$-orientation of $E$ cannot be carried onto $E'$ due to the possibility that $\phi(G_1)$ or $\phi(G_2)$ in $E'[A]$ has an order smaller than $A$. To address this, the degree of the isogeny must be relatively prime to $A$. The following definition formalizes this notion:

\begin{definition}
    For two artificially $A$-oriented curves $(E, \frak{A})$ and $(E', \frak{A}')$, and an integer $B$ relatively prime to the $A$, the pairs is said to be $B$-isogenous if there exists a $B$-isogeny $\phi:E\rightarrow E'$ such that $$\frak{A}'= (G'_{1}, G'_{2}) =\phi(G_1, G_2 )=\phi(\frak{A}).$$
\end{definition}

With fixed generators $\langle P_1 \rangle = G_1$ and $\langle P_2 \rangle = G_2$, the subgroups $G'_1$ and $G'_2$ are represented as $[\alpha]\phi(P_1)$ and $[\beta]\phi(P_2)$, respectively, for $\alpha, \beta \in \mathbb{Z}/A\mathbb{Z}$. Although artificial orientations do not generate a commutative group action as in standard orientations \cite{osidh}, they provide sufficient structure for computing parallel isogenies. Concretely, given $A$-oriented curves $(E, \mathfrak{A})$ and $(E', \mathfrak{A}')$, connected by a $\mathfrak{B}$-isogeny $\phi: E \to E'$, where $\mathfrak{A} = (G_1, G_2)$ and $\mathfrak{A}' = (G'_1, G'_2)$, the isogenies $\psi_1 : E \to E_1$ and $\psi_2 : E' \to E_2$ are parallel, as depicted in Figure \ref{fig:Parallel Isogenies}.
\begin{figure}
    \centering
    \begin{tikzcd}[row sep=scriptsize, column sep=huge]
    E\arrow[r, "\phi"]\arrow[d,swap,"\psi_1"]&E^{'}\arrow[d,"\psi_2"]\\ 
    E_1&E_2
\end{tikzcd}
    \caption{Parallel Isogenies}
    \label{fig:Parallel Isogenies}
\end{figure}
Here, $E_1 = E / \langle [A_1]G_1 + [A_2]G_2 \rangle$ and $E_2 = E' / \langle [A_1]G'_1 + [A_2]G'_2 \rangle$, with $\ker(\psi_2) = \phi(\ker(\psi_1))$. The codomain curves $E_1$ and $E_2$ are $B$-isogenous, connected by the isogeny $\phi'$ with $\ker(\phi') = \psi_1(\ker(\phi))$. The isogenies $\psi_1$ and $\psi_2$ are thus characterized by the multiplicative decomposition $A = A_1 A_2$. The properties of artificial orientation are leveraged in the pre-signature and adaptation phases of our scheme.

\subsection{Computational Hardness Assumptions}

The following computational hardness assumptions, which are derived from the generic problem of finding an isogeny between two isogenous elliptic curves defined over a field $k$, are presumed to be computationally infeasible. These assumptions underpin the security of our scheme and are employed throughout its construction.

\begin{problem}[Supersingular Smooth Endomorphism Problem \cite{sqisign}] Given a prime $p$ and a supersingular elliptic curve $E/\mathbb{F}_{p^2}$, find a (non-trivial) cyclic endomorphism of $E$ of smooth degree.
\label{problem:ERP}
\end{problem}

\begin{problem}[SSIP-A \cite{artificial}] \label{problem:SSIP-A}
    Let $(E,\frak{B})$ be an artificially $B$-oriented curve, and let $A$ be an integer coprime to $B$. Let $\phi: E \rightarrow E'$ be a cyclic isogeny of degree $A$ and let $\frak{B}' = \phi(\frak{B})$. Given $(E, \frak{B})$ and $(E', \frak{B}')$ and the degree $A$, compute $\phi$.
\end{problem}

\begin{problem}[SSIP-B \cite{artificial}] \label{problem:SSIP-B}
    Let $(E, \frak{B})$ be an artificially $B$-oriented curve, and let $A$ be an integer coprime to $B$. Let $\psi : E \rightarrow E'$ be a cyclic $\frak{B}$-isogeny of degree $B$, with $A < B$. Let $P$, $Q$ be a basis of $E[A]$. Given $(E, \frak{B})$, the points $P,Q$, and the curve $E'$ with the points $\psi(P)$ and $\psi(Q)$, compute $\psi$.
\end{problem}

\subsection{Adaptor Signature Scheme}

We begin by recalling the definition of a cryptographically hard relation:

\begin{definition}[Hard Relation]
    Let $\mathsf{R\subseteq W\times S}$ be a set of witness/statement pairs $\mathsf{(w, s)}$. The language of $\mathsf{R}$ is defined as: $\mathcal{L}_\mathsf{R}:= \{\mathsf{s \mid \exists\, w \text{  s.t.  } \mathsf{(w, s)} \in R}\}$. The relation $\mathsf{R}$ is said to be a \textit{hard relation} if the following conditions are satisfied:
    \begin{itemize}
        \item[-] There exists a $\mathsf{PPT}$ algorithm $\mathsf{GenR(1^\lambda)}$ taking the security parameter $\lambda$ as input, and outputs a witness/statement pair $\mathsf{(w, s) \in R}$.
        \item[-] The relation's validation is decidable in polynomial time.
        \item[-] For any $\mathsf{PPT}$ adversary $\mathcal{A}$, a negligible function $\mathsf{negl}$ exists such that:
    \[ \text{Pr}\left[\left.  
    \begin{array}{cc}
    \\
    \mathsf{(w^{\ast},s)\in R}\\
    \\
    \end{array} \right| \begin{array}{cc}
    \mathsf{(w,s)\leftarrow GenR(1^\lambda)} \\
    \mathsf{w^\ast \leftarrow \mathcal{A}(s)}
    \end{array} \right]\leq \mathsf{Negl(\lambda)}.\]
    \end{itemize}
\end{definition}

\noindent\textbf{Non-interactive Proof System.} Let $\mathsf{(w,s)\in R}$ be cryptographically a hard relation, and $\mathcal{H}$ be a random oracle. A \textit{non-interactive proof system} is a pair of $\mathsf{PPT}$ oracle algorithms $\mathsf{(P,V)}$, where: 
\begin{description}
    \item[-] $\mathsf{\pi_w/\perp\leftarrow P^{\mathcal{H}}(w, s)}$: A prover $\mathsf{P}$ takes a pair $\mathsf{(w, s)\in R}$ as input and outputs a proof $\mathsf{\pi_{w}}$ of the statement $\mathsf{s}$ with witness $\mathsf{w}$. If $\mathsf{(w, s)\not\in R}$, $\mathsf{ P^{\mathcal{H}}(w, s)=\perp}$.
    \item[-] $\mathsf{0/1\leftarrow V^{\mathcal{H}}(s, \pi_w)}$: A verifier $\mathsf{V}$ takes a pair $\mathsf{(s, \pi_{w})}$ and outputs whether the proof $\mathsf{\pi_w}$ for $\mathsf{s}$ is valid.
\end{description} 
This system satisfies the following conditions:
\begin{itemize}
    \item[i.] Completeness: If $\mathsf{(w,s)\in R}$ and $\mathsf{\pi_w\leftarrow P^{\mathcal{H}}(w, s)}$, then there exists a negligible function $\mathsf{negl}$ such that $\text{Pr}[\mathsf{V^{\mathcal{H}} =1]\geq 1-negl(\lambda)}$.
    \item[ii.] Zero-knowledge ($\mathsf{NIZK}$): For a $\mathsf{PPT}$ algorithm $\mathcal{S}$, any $\mathsf{(w,s)}$, and a $\mathsf{PPT}$ algorithm $\mathcal{D}$, the following distributions are computationally indistinguishable:
    \begin{description}
        \item[-]$\mathsf{\pi_w\leftarrow P^{\mathcal{H}}(w, s)}$ if $\mathsf{(w, s)\in R}$ and $\mathsf{\pi_{w}\leftarrow\perp}$ otherwise. Output $\mathcal{D^H}\mathsf{(w,s,\pi_{w})}$.
        \item[-] $\mathsf{\pi_w\leftarrow \mathcal{S}(s, 1)}$ if $\mathsf{(w, s)\in R}$ and $\mathsf{\pi_{w}\leftarrow\mathcal{S}(s,0)}$ otherwise. Output $\mathcal{D^H}\mathsf{(w,s,\pi_{w})}$.
    \end{description}
    \item[iii.]  Online-extractability: For a $\mathsf{PPT}$ algorithm $\mathcal{E}$, and any algorithm $A$, let $(\mathsf{s},\pi_{\mathsf{w}})\leftarrow A^{\mathcal{H}}(\lambda)$ be the sequence of queries of $A$ to $\mathcal{H}$, and $H_A$ be the $\mathcal{H}$’s answers. Let $\mathsf{w} \leftarrow \mathcal{E}(s,\pi_w ,H_A)$. Then it holds that
    $$\text{Pr}[\mathsf{(w,s)\not\in R \land V^{\mathcal{H}}(s,\pi_{w})=1]\leq negl(\lambda)}.$$
\end{itemize}

\noindent\textbf{Digital Signature Scheme.} We recall the definition of a digital signature scheme and the properties that a signature scheme must satisfy in order to be considered secure.

\begin{definition}[Digitial Signature Scheme] 
A digital signature is a triple $\mathsf{\Sigma = (KeyGen, Sig, Ver)}$ consisting of three polynomial-time algorithms:
\begin{itemize}
   \item[-] ${\mathsf{(sk, pk)}\leftarrow \mathsf{KeyGen}(1^\lambda):}$ a $\mathsf{PPT}$ algorithm that takes security parameter $\lambda$ as input, outputs a secret/public key pair $\mathsf{(sk, pk)}$.
   \item[-] ${\mathsf{\sigma}\leftarrow \mathsf{Sig}(\mathsf{sk}, m):}$ a $\mathsf{PPT}$ algorithm that takes a secret key $\mathsf{sk}$ and a message $m \in \{0, 1\}^{\ast}$ as input, outputs a signature $\sigma$ for the message $m$.
   \item[-] ${\mathsf{0/1}\leftarrow\mathsf{Ver}(\mathsf{pk},m,\sigma):}$ a $\mathsf{DPT}$ algorithm that takes a public key $\mathsf{pk}$, a message $m \in \{0, 1\}^{\ast}$, and signature $\sigma$ as input, outputs a bit $b\in\{0,1\}$.
\end{itemize}
\end{definition}

A signature scheme is \emph{correct} if, for any security parameter $\lambda\in \mathbb{N}$, any key pair $(\mathsf{sk,pk})\leftarrow \mathsf{KeyGen(1^\lambda)}$, and for any message $m\in\{0,1\}^\ast$, the following holds: $$\text{Pr}\Big[\mathsf{Ver}(\mathsf{pk},m,\mathsf{Sig}(\mathsf{sk},m))=1\, \Big|\, (\mathsf{sk,pk})\leftarrow \mathsf{KeyGen(1^\lambda)}\Big]=1.$$

There are several security requirements for a signature scheme, one of the most common being \textit{existential unforgeability under chosen message attacks} (\textsf{EUF-CMA}). This property ensures that forging a verifiable signature on a message $m$  without knowledge of the secret key $\mathsf{sk}$ is infeasible, even if the \textsf{PPT} adversary has access to many valid signatures on messages of its choice but message $m$. The formal definition of this property is as follows:

\begin{definition}[$\mathsf{EUF}$-$\mathsf{CMA}$ Security] A signature scheme $\mathsf{\Sigma}
$ is $\mathsf{EUF}$-$\mathsf{CMA}$ secure if for every $\mathsf{PPT}$ adversary $\mathcal{A}$, there exists a negligible function $\mathsf{negl}$ such that 
$$\text{Pr} [\mathsf{SigForge}_{\mathcal{A},\mathsf{\Sigma}}(\lambda)=1]\leq \mathsf{negl}(\lambda),$$
where the experiment $\mathsf{SigForge}_{\mathcal{A},\mathsf{\Sigma}}$ is defined as follows:

\begin{pchstack}[boxed, center, space=2em]
    
    \procedure[linenumbering]{$\mathsf{SigForge}_{\mathcal{A},\mathsf{\Sigma}}(\lambda)$}{%
    \mathcal{Q}\leftarrow \emptyset\\
    (\sk, \pk)\leftarrow \mathsf{KeyGen}(\secparam)\\
    (m,\sigma)\leftarrow \adv^{\mathcal{O}_S}(\pk)\\
    \pcreturn (m\not\in\mathcal{Q}\land \mathsf{Ver}(\pk,m,\sigma))
    }
    
    \procedure[linenumbering]{$\mathcal{O}_S (m)$}{%
    \sigma \leftarrow \mathsf{Sig}(\sk,m)\\ 
    \mathcal{Q}:=\mathcal{Q}\cup\{m\}\\
    \pcreturn \sigma
    }
\end{pchstack}
\end{definition}

A stronger definition is \textit{strong existential unforgeability under chosen message attacks} (\textsf{SUF-CMA}), which ensures the difficulty of transforming a valid signature on a message $m$ into another valid signature on $m$. %Specifically, in case an adversary is given a message/signature pair $(m, \sigma)$ along with some chosen message/signature pairs, the signature scheme is considered \textsf{SUF-CMA} secure if the adversary cannot produce a new signature $\sigma^\ast$ for the message $m$. 
The formal definition is as follows:

\begin{definition}[$\mathsf{SUF}$-$\mathsf{CMA}$ Security] \label{def:StrongSigForge} A signature scheme $\mathsf{\Sigma}$ is $\mathsf{SUF}$-$\mathsf{CMA}$ secure if for every $\mathsf{PPT}$ adversary $\mathcal{A}$, there exists a negligible function $\mathsf{negl}$ such that 
$$\text{Pr} [\mathsf{StrongSigForge}_{\mathcal{A},\mathsf{\Sigma}}(\lambda)=1]\leq \mathsf{negl}(\lambda),$$
where the experiment $\mathsf{StrongSigForge}_{\mathcal{A},\mathsf{\Sigma}}$ is defined as follows:

\begin{pchstack}[boxed, center, space=2em]
    
    \procedure[linenumbering]{$\mathsf{StrongSigForge}_{\mathcal{A},\mathsf{\Sigma}}(\lambda)$}{%
    \mathcal{Q}\leftarrow \emptyset\\
    (\sk, \pk)\leftarrow \mathsf{KeyGen}(\secparam)\\
    (m,\sigma)\leftarrow \adv^{\mathcal{O}_S}(\pk)\\
    \pcreturn ((m,\sigma)\not\in\mathcal{Q}\land \mathsf{Ver}(\pk,m,\sigma))
    }
    
    \procedure[linenumbering]{$\mathcal{O}_S (m)$}{%
    \sigma \leftarrow \mathsf{Sig}(\sk,m)\\ 
    \mathcal{Q}:=\mathcal{Q}\cup\{m,\sigma\}\\
    \pcreturn \sigma
    }
\end{pchstack}    
\end{definition}

\noindent\textbf{Adaptor Signature Scheme.} An adaptor signature is a cryptographic primitive that extends an ordinary digital signature. It hides secret randomness within the signature, which is only revealed once the signature is generated. The process begins with the generation of a pre-signature, which is then adapted into a full signature by applying secret randomness. In the final step, this secret randomness is extracted through cryptographic procedures. The signature produced is verifiable using the verification algorithm of the underlying signature scheme. An adaptor signature also has specific security properties. For any statement $\mathsf{s} \in \mathcal{L}_\mathsf{R}$, a signer with secret key \textsf{sk} can produce a pre-signature $\tilde{\mathsf{\sigma}}$ on any message $m$. This pre-signature can be adapted into a full signature $\mathsf{\sigma}$ if and only if the user has a witness \textsf{w} to the statement \textsf{s}. Additionally, anyone with access to the pre-signature $\tilde{\mathsf{\sigma}}$, (full) signature $\mathsf{\sigma}$, and statement \textsf{s} can extract the witness \textsf{w}, thus revealing the hard relation.

The formal definition of an adaptor signature scheme and its properties are given as follows:

\begin{definition}[Adaptor Signature Scheme]
An adaptor signature scheme with respect to a hard relation $\mathsf{R}$ and a signature scheme $\mathsf{\Sigma} = ( \mathsf{KeyGen}, \mathsf{Sig}, \mathsf{Ver})$ is a quadruple  $\Xi_{\mathsf{R,\Sigma}}=(\mathsf{PreSig}, \mathsf{PreVer}, \mathsf{Adapt},\mathsf{Ext})$ defined as:

\begin{itemize}
    \item[-] $\tilde{\sigma}\leftarrow\mathsf{PreSig}(\mathsf{sk}, m, \mathsf{s}):$ a $\mathsf{PPT}$ algorithm that takes a secret key $\mathsf{sk}$, a message $m \in \{0,1\}^\ast$, and a statement $\mathsf{s} \in \mathcal{L}_\mathsf{R}$, outputs a pre-signature $\Tilde{\sigma}$.
    \item[-] $0/1\leftarrow\mathsf{PreVer}(\mathsf{pk}, m, \mathsf{s}, \Tilde{\sigma}):$ a $\mathsf{DPT}$ algorithm that takes a public key $\mathsf{pk}$,  a message $m \in \{0,1\}^\ast$, a statement $\mathsf{s} \in \mathcal{L}_\mathsf{R}$, and a pre-signature $\Tilde{\sigma}$, produces a bit $b\in\{0,1\}$.
    \item[-] $\sigma\leftarrow\mathsf{Adapt(\Tilde{\sigma},w)}:$ a $\mathsf{DPT}$ algorithm that takes a valid pre-signature $\Tilde{\sigma}$, and a witness $\mathsf{w}$, generates a signature $\sigma$.
    \item[-] $\mathsf{w/\perp \leftarrow Ext(\sigma,\Tilde{\sigma},s}):$ a $\mathsf{DPT}$ algorithm that takes a  pre-signature $\Tilde{\sigma}$, a corresponding signature $\sigma$, and a statement $\mathsf{s} \in \mathcal{L}_\mathsf{R}$, produces a witness $\mathsf{w}$ to the statement $\mathsf{s}$, or $\perp$.    
\end{itemize}
\end{definition}

In an adaptor signature scheme, $\Xi_{\mathsf{R,\Sigma}}$, the algorithm $\mathsf{GenR}$ generates witness/statement pairs $\mathsf{(w, s)}$ based on the underlying hard relation $\mathsf{R}$. As mentioned earlier, several properties ensure the security of an adaptor signature scheme. The first property is \textit{pre-signature correctness}, which guarantees that an honestly generated pre-signature can be adapted into a valid signature.

\begin{definition}[Pre-signature Correctness] An adaptor signature scheme $\Xi_{\mathsf{R,\Sigma}}$ satisfies pre-signature correctness if for any $\lambda\in \mathbb{N}$, any message $m\in\{0,1\}^\ast$, and any witness/statement pair $\mathsf{(w,s)}$, the following holds:
\[ \text{Pr}\left[\left.  
    \begin{array}{cc}
    \mathsf{PreVer}(\mathsf{pk},m,\mathsf{s},\tilde{\sigma})=1\\
    \mathsf{Ver}(\mathsf{pk},m,\sigma)=1   \\
    \mathsf{(w',s)\in R}
    \end{array} \right. \left| \begin{array}{cc}
    
    \mathsf{(sk,pk)}\leftarrow \mathsf{KeyGen(1^\lambda)} \\
    \tilde{\sigma}\leftarrow \mathsf{PreSig(sk},m,\mathsf{s})\\
    \sigma := \mathsf{Adapt(\tilde{\sigma},w)}\\
    \mathsf{w' := Ext(\sigma,\tilde{\sigma},s)}
    \end{array}  \right.\right]=1.\]
    \label{def:correctness}
\end{definition}

The second property of an adaptor signature is \textit{pre-signature adaptability}. It states that any valid (though not necessarily honestly generated) pre-signature for a statement \textsf{s} can be adapted into a valid signature using a witness \textsf{w} such that $\mathsf{(w, s) \in R}$.

\begin{definition}[Pre-signature Adaptability] An adaptor signature scheme $\Xi_{\mathsf{R,\Sigma}}$ satisfies pre-signature adaptability if for any $\lambda\in\mathbb{N}$,  message $m\in \{0,1\}^\ast$, witness/statement pair $\mathsf{(w,s)\in R}$, key pair $\mathsf{(sk,pk)}\leftarrow \mathsf{KeyGen}(1^\lambda)$, and pre-signature $\tilde{\sigma}\leftarrow \{0,1\}^\ast$ such that $\mathsf{PreVer(pk},m,\mathsf{s},\tilde{\sigma})=1$, the following holds: 
$$\text{Pr}[\mathsf{Ver(pk},m,\mathsf{Adapt(\tilde{\sigma},w)})=1]=1.$$
\end{definition}

Another key property is \textit{existential unforgeability under chosen message attack} (\textsf{aEUF-CMA}). This property states that even with access to a pre-signature on a message $m$ with respect to a random statement $\mathsf{s} \in \mathcal{L}_\mathsf{R}$, it is computationally infeasible for an adversary to forge a valid signature $\mathsf{\sigma}$ for $m$.

\begin{definition}[$\mathsf{aEUF}\text{-}\mathsf{CMA}$ Security]\label{def:aEUF-CMA} An adaptor signature scheme $\Xi_{\mathsf{R,\Sigma}}$ is $\mathsf{aEUF}\text{-}\mathsf{CMA}$ secure if for any $\mathsf{PPT}$ adversary $\mathcal{A}$, there exists a negligible function $\mathsf{negl}$ such that 
$$\text{Pr}[\mathsf{aSigForge}_{\mathcal{A},\Xi_{\mathsf{R,\Sigma}}}(\lambda)=1]\leq \mathsf{negl}(\lambda),$$
where the experiment $\mathsf{aSigForge}_{\mathcal{A},\Xi_{\mathsf{R,\Sigma}}}$ is defined as follows:

\begin{pchstack}[boxed, center, space=2em]
    \procedure[linenumbering]{$\mathsf{aSigForge}_{\mathcal{A},\Xi_{\mathsf{R,\Sigma}}}(\lambda)$}{%
    \mathcal{Q}:=\emptyset \\
    (\sk,\pk)\leftarrow \mathsf{KeyGen}(\secparam)\\
    m\leftarrow \adv^{\mathcal{O}_S ,\mathcal{O}_{pS}}(\pk)\\
    \mathsf{(w,s)}\leftarrow \mathsf{GenR(1^{\lambda})}\\
    \tilde{\sigma}\leftarrow\mathsf{PreSig}(\sk,m,\mathsf{s})\\
    \sigma\leftarrow\adv^{\mathcal{O}_S ,\mathcal{O}_{pS}}(\tilde{\sigma},\mathsf{s})\\
    \pcreturn m\not\in\mathcal{Q}\land\mathsf{Ver}(\pk,m,\sigma)
    }

    \begin{pcvstack}[space=1em]
    
    \procedure[linenumbering]{$\mathcal{O}_S (m)$}{%
    \sigma\leftarrow \mathsf{Sig}(\sk,m)\\
    \mathcal{Q}:=\mathcal{Q}\cup\{m\}\\
    \pcreturn \sigma
    }
    \procedure[linenumbering]{$\mathcal{O}_{pS} (m,s)$}{%
    \tilde{\sigma}\leftarrow\mathsf{PreSig}(\sk,m,\mathsf{s})\\
    \mathcal{Q}:=\mathcal{Q}\cup\{m\}\\
    \pcreturn \tilde{\sigma}
    }
    \end{pcvstack}
\end{pchstack}

\end{definition}

The fourth and last property is called \textit{witness extractability}. This property guarantees that once a pre-signature is adapted into a (full) signature, it must not be the case that the witness for the original statement used to generate the pre-signature cannot be extracted.

\begin{definition}[Witness Extractability] \label{def:aWitExtGame} An adaptor signature scheme $\Xi_{\mathsf{R,\Sigma}}$ is witness extractable if for any $\mathsf{PPT}$ adversary $\mathcal{A}$, there exists a negligible function $\mathsf{negl}$ such that the following holds:
$$\text{Pr}[\mathsf{aWitExt}_{\mathcal{A},\Xi_{\mathsf{R,\Sigma}}}(\lambda)=1]\leq \mathsf{negl}(\lambda),$$
where the experiment $\mathsf{aWitExt}_{\mathcal{A},\Xi_{\mathsf{R,\Sigma}}}$ is defined as follows:

\begin{pchstack}[boxed, center, space=2em]
    
    \procedure[linenumbering]{$\mathsf{aWitExt}_{\mathcal{A},\Xi_{\mathsf{R,\Sigma}}}(\lambda)$}{%
    \mathcal{Q}:=\emptyset \\
    (\sk,\pk)\leftarrow \mathsf{KeyGen}(\secparam)\\
    (m,\mathsf{s})\leftarrow \adv^{\mathcal{O}_S,\mathcal{O}_{pS}}(\pk)\\
    \tilde{\sigma}\leftarrow \mathsf{PreSig}(\sk,m,\mathsf{s})\\
    \sigma\leftarrow\adv^{\mathcal{O}_S ,\mathcal{O}_{pS}}(\tilde{\sigma})\\
    \mathsf{w}^{\ast}:=\mathsf{Ext}(\sigma,\tilde{\sigma},\mathsf{s})\\
    \pcreturn (m\not\in\mathcal{Q}\land (\mathsf{w}^{\ast} ,\mathsf{s})\not\in \mathsf{R}\land \mathsf{Ver}(\pk,m,\sigma))
    }

    \begin{pcvstack}[space=1em]
    
    \procedure[linenumbering]{$\mathcal{O}_S (m)$}{%
    \sigma\leftarrow \mathsf{Sig}(\sk,m)\\
    \mathcal{Q}:=\mathcal{Q}\cup\{m\}\\
    \pcreturn \sigma
    }
    \procedure[linenumbering]{$\mathcal{O}_{pS} (m,\mathsf{s})$}{%
    \tilde{\sigma}\leftarrow\mathsf{PreSig}(\sk,m,\mathsf{s})\\
    \mathcal{Q}:=\mathcal{Q}\cup\{m\}\\
    \pcreturn \tilde{\sigma}
    }
    \end{pcvstack}
\end{pchstack}
\end{definition}

In light of the above properties of the adaptor signature scheme, the following definition is established: 

\begin{definition}[Secure Adaptor Signature Scheme] An adaptor signature scheme $\Xi_{\mathsf{R,\Sigma}}$ is secure if it is $\mathsf{aEUF}$-$\mathsf{CMA}$ secure, pre-signature adaptable, and witness extractable.
\end{definition}

\subsection{SQIsignHD}

SQIsignHD \cite{sqisignhd} is a post-quantum digital signature scheme derived from SQISign \cite{sqisign}, incorporating recent advancements stemming from attacks \cite{CDattackSIDH, attack2SIDH, attack3} on SIDH. These advancements enable efficient representation of isogenies of arbitrary degrees. In comparison to SQISign, SQIsignHD provides improved scalability for higher security levels, greater simplicity and efficiency, and smaller signature sizes. The protocol is outlined as follows:

Let $D_\varphi := \prod_{i=1}^{n}\ell_{i}^{e_i}$ be a smooth integer and $\mu(D_\varphi):=\prod_{i=0}^n \ell_i^{e_i-1}(\ell_i + 1)$. Also, let $\mathit{\Phi}_{D_\varphi}(E, h)$ be an arbitrary function that maps an integer $h\in[1, \mu(D_\varphi)]$ to a
non-backtracking isogeny of degree $D_\varphi$ starting at $E$. Consider a hash function $\mathsf{H} : \{0, 1\}^{\ast} \to [1, \mu(D_\varphi) ]$ which is cryptographically secure.
%\begin{figure}[h]
%\centering
%\begin{tikzcd}[column sep=40pt]
%E_0 \arrow[r, "\psi", dashed] \arrow[d, swap, "\tau", dashed]
%& E_1 \arrow[bend right=20,swap]{d}{\sigma} \\ [20pt]
%E_A \arrow[r, swap, "\varphi"]
%& E_2
%\end{tikzcd}
%\caption{SQIsignHD Protocol}
%\label{fig:sqisign}
%\end{figure}
\begin{description}
    \item[$\mathsf{Setup}.$] Choose a prime $p$ and supersingular elliptic curve $E_0/\mathbb{F}_{p^2}$ with known endomorphism ring $\mathcal{O}_0\cong\groupendo(E_0)$, where $E_0$ has smooth torsion defined over a small extension of $\mathbb{F}_{p^2}$ of degree 1 or 2.
    \item[$\mathsf{KeyGen}.$] Generate a random secret isogeny $\tau : E_0 \to E_A$ of fixed smooth degree $D_\tau$. The secret/public key pair is $\mathsf{(sk,pk)}:=(\tau, E_A)$.
    \item[$\mathsf{Sign}.$] Generate a random (secret) commitment isogeny $\psi : E_0 \to E_1$. For signing a message $m$, build the isogeny $\mathit{\Phi}_{D_\varphi}(E_{A},h)= \varphi: E_A \to E_2$, where $h = \mathsf{H} (j(E_1), m)$. From the knowledge of the secret key $\tau$, and isogenies $\varphi, \psi$, construct an efficient representation $R =(\sigma(P_1),\sigma(P_2),q)$ given by the image of torsion points by a response isogeny $\sigma: E_1 \to E_2$ and return the pair $\mathsf{\Sigma}:=(E_1, R)$ as a signature.
    \item[$\mathsf{Verify}.$] Upon receiving a signature $\mathsf{\Sigma}=(E_1, R)$ associated with the message $m$ and public key $E_A$, the verifier recovers $h = H(j(E_1 ),m)$ and then computes $\varphi = \Phi(E_{A},h): E_A \to E_2$. Finally, the verifier checks that $R$ represents correctly an isogeny $\sigma: E_1 \to E_2$ by computing a higher dimensional isogeny, as described in SQIsignHD.
\end{description}

 The public parameters for SQIsignHD are easy to generate. Specifically, the underlying prime is of the form $p = c\ell^{f}\ell^{'f'} - 1$, where $\ell$ and $\ell'$ are distinct primes (in practice, $\ell = 2$ and $\ell' = 3$), $c \in \mathbb{N}^{\ast}$ is a small cofactor, and $\ell^{f} \approx \ell^{'f'} \approx p^{1/2}$. This ensures sufficient accessible torsion for isogeny computations. This flexibility allows replacing $\ell^{f}$ and $\ell^{'f'}$ with a collection of small primes, as discussed in Section \ref{sec3.1}, providing a suitable setting for applying artificial orientation in our construction.
 
The signature, as shown in the protocol, is the data $(E_1,\sigma(P_1),\sigma(P_2),q)$, with $q \approx p^{1/2}$, $\sigma: E_1\to E_2$ a $q$-isogeny, and $(P_1, P_2)$ a basis of $E_1 [\ell^{f}]$. This data is based on the following definition:

\begin{definition}[\cite{sqisignhd}]
    Suppose that $\mathsf{A}$ is an algorithm and $\varphi: E\to E^{'}$ is an $\mathbb{F}_q$-rational isogeny. Then, an \textit{efficient representation} of isogeny $\varphi$ (with respect to $\mathsf{A}$) is some data $\mathsf{D}\in\{0,1\}^{\ast}$ such that:
    \begin{itemize}\setlength\itemsep{0.05em}
        \item[1.] $\mathsf{D}$ has polynomial size in $\log(\deg(\varphi))$ and $\log(q)$.
        \item[2.] On input $\mathsf{D}$ and $P\in E(\mathbb{F}_{q^k} )$, $\mathsf{A}$ returns $\varphi(P)$ in polynomial time in $k\log(q)$ and $\log(\deg(\varphi))$.
    \end{itemize}
    \label{def:efficientRepresentation}
\end{definition}

\section{New Adaptor Signature Construction}\label{sec3}
In this section, we present a new post-quantum adaptor signature scheme built upon SQIsignHD \cite{sqisignhd} as the underlying signature scheme. To incorporate the associated hard relation, we utilize the hybrid variant of \textsf{bin}SIDH, denoted as $\textsf{bin}\text{SIDH}^{\textsf{hyb}}$, introduced in Section 5 of \cite{artificial}. This variant combines oriented and non-oriented approaches, wherein one party computes \textsf{bin}SIDH-like isogenies while the other performs SIDH-like isogenies.

Currently, the only secure post-quantum isogeny-based adaptor signature scheme is IAS, proposed in \cite{ias}, which is built upon CSI-FiSh \cite{csifish}. However, IAS faces efficiency limitations due to the parameter sizes required by CSI-FiSh. Specifically, CSI-FiSh operates at most on the CSIDH-512 parameters, as efficient computation of the class group action on uniformly random group elements necessitates prior knowledge of the class group structure. In the following, we provide a detailed description of our proposed post-quantum adaptor signature scheme and present the corresponding protocol in Algorithm \ref{alg:SQIAsignHD}.

\subsection{Public Parameters}\label{sec3.1}  
To deploy our protocol, we first establish a set of initial public parameters. These parameters are inspired by those employed in $\mathsf{bin}$SIDH$^\mathsf{hyb}$ and SQIsignHD. The setup of our scheme is defined as follows.  

We select a prime $p$ of the form $p = ABCf - 1$, where $A = 2^a$, $B = \prod_{i=1}^{t} \ell_i$, and $C = 3^c$ are pairwise relatively prime integers. Here, $f$ is a small cofactor, $\ell_i$'s represent distinct small primes, and the sizes of $A$, $B$, and $C$ are chosen such that $A \approx C \approx p^{1/4}$ and $B \approx p^{1/2}$. Let $E_0 / \mathbb{F}_{p^2}$ denote a supersingular elliptic curve with a known endomorphism ring $\groupendo(E_0) \cong \mathcal{O}_0 \subset \mathcal{B}_{p,\infty}$, and assume $|E_0(\mathbb{F}_{p^2})| = (p+1)^2$. We define $\mathfrak{B} = (G_1, G_2)$ as an artificial $B$-orientation on $E_0$, and fix a basis $\langle P, Q \rangle = E_0[C]$. Additionally, we employ a cryptographically secure hash function $\mathsf{H} : \{0, 1\}^* \to [1, \mu(D_\varphi)]$, analogous to the one used in SQIsignHD.  

\subsection{Key Generation \& Hard Relation}  
The key generation procedure follows the standard process in SQIsignHD. Specifically, a random secret isogeny $\tau: E_0 \rightarrow E_\tau$ is generated, and the secret/public key pair is defined as $\mathsf{(sk, pk)} := (\tau, E_\tau)$.  

To define the hard relation in our scheme, we set the witness/statement pairs as follows:  
\[ \mathsf{R}_{\frak{A}}:=\left\{\left.  
    \begin{array}{cc}
    \left(w,I_w : =(E_w , w(\frak{B}),\pi_w)\right)
    \end{array} \right. \left| \begin{array}{cc}
    w:E_0 \rightarrow E_w := E_0 / \langle  P + [\alpha] Q\rangle,\\ 
    \text{where }\langle P,Q\rangle = E_0 [C], \alpha\in \mathbb{Z}/C\mathbb{Z}.\\
    (E_0, \frak{B}) \text{ is artificially $B$-oriented.}
    \end{array}  \right.\right\},\]

\noindent where $w$ denotes the secret witness isogeny with the artificially $B$-oriented curve $(E_0, \mathfrak{B})$ as its domain, while $(E_w, w(\mathfrak{B}))$ constitutes the statement, consisting of the target elliptic curve $E_w$ and the image of the artificial $B$-orientation $\mathfrak{B} = (G_1, G_2)$ under the isogeny $w$. Additionally, $\pi_w$ denotes a zero-knowledge proof that $(w, (E_w, w(\mathfrak{B})))$ is a valid instance of the hard relation $\mathsf{R}_{\mathfrak{A}}$. 

\subsection{Pre-signature} 

The pre-signing algorithm shares similarities with the signing procedure described in the SQIsignHD protocol but introduces notable differences, particularly in generating the commitment isogeny (and the corresponding curve) and incorporating additional elements required during the adaptation phase.

Unlike SQIsignHD, our scheme’s pre-signature phase involves two (secret) commitment isogenies. The first serves a role similar to the commitment isogeny in SQIsignHD, while the second, generated in conjunction with the statement curve, lays the foundation for the adaptation phase. We now examine these components in detail.\\

\noindent\textbf{Commitment \pmb{$\psi$}.}  
The first commitment isogeny, $\psi$, is a $\mathfrak{B}$-oriented isogeny $\psi : E_0 \rightarrow E_\psi$, generated by uniformly sampling a vector $\vec{b}$ from $\{1, 2\}^t$ to compute  
\[
\ker(\psi) := \langle G_{b_1}^1, G_{b_2}^2, \ldots, G_{b_t}^t \rangle,  
\]  
where \( G_1 := \langle G_1^1, G_1^2, \ldots, G_1^t \rangle \) and \( G_2 := \langle G_2^1, G_2^2, \ldots, G_2^t \rangle \), with \( |G_1^i| = |G_2^i| = \ell_i \), for \( 1 \leq i \leq t \).  

Furthermore, using the isogeny $\psi$, we compute the images of the publicly given points $P$ and $Q$ under $\psi$. These images are denoted as $S := (\psi(P), \psi(Q))$.\\

\noindent\textbf{Commitment \pmb{$\psi’$}.}
After parsing $I_w$ as  $(E_w,w(\mathfrak{B}),\pi_w)$ and verifying that $1=\mathsf{NIZK.V}(E_w,\pi_{w})$, the second commitment isogeny, $\psi'$, is derived by pushing forward the first commitment isogeny, $\psi$, through the witness $w: E_0 \to E_w$ using the component $w(\mathfrak{B})$ of the public statement. Here, $w(\mathfrak{B})$ represents the image of the artificially $B$-orientation $\mathfrak{B}$ under the witness isogeny $w$. Formally, this is defined as $\psi' := [w]_{\ast}\psi : E_w \to E_1$.

As a result, we obtain the second commitment curve, $E_1$, whose $j$-invariant is used to compute the challenge isogeny. Finally, we compute a zero-knowledge proof, $\pi_{\psi'}$, to demonstrate that $E_1$ is the codomain of the isogeny parallel to $\psi$.\\

Now, the challenge and pre-signature isogenies are constructed as follows:  \\

\noindent\textbf{Challenge \pmb{$\varphi$}.}  
To generate the challenge isogeny, the $j$-invariant of the second commitment curve $E_1$ is combined with a message $m$ to produce an isogeny starting at the public key $E_\tau$. Specifically, for $h := \mathsf{H}(j(E_1), m)$, the challenge isogeny is defined as $\varphi := \Phi(E_\tau, h): E_\tau \to E_2$.

\noindent\textbf{Pre-signature \pmb{$\tilde{\sigma}$}.}  
To complete the pre-signing phase for a message $m$, given knowledge of the endomorphism ring $\text{End}(E_0) \cong \mathcal{O}_0$ and the isogenies $\tau$, $\varphi$, and $\psi$, an efficient representation $\mathcal{R}_{\tilde{\sigma}} := (\tilde{\sigma}(R_1), \tilde{\sigma}(R_2), \deg(\tilde{\sigma}))$ is constructed. This representation is derived from the images of a canonically determined basis $\langle R_1, R_2 \rangle$ of $E_\psi[A]$ under the pre-signature isogeny $\tilde{\sigma}: E_\psi \to E_2$.\\

Thus, the pre-signature tuple is defined as $
\tilde{\mathsf{\Sigma}} := (E_1, \pi_{\psi'}, E_\psi, S, \mathcal{R}_{\tilde{\sigma}})$,  
and the pre-signing algorithm is formally expressed as  
$$\tilde{\mathsf{\Sigma}} \leftarrow \mathsf{PreSig}(\mathsf{sk}, m, \mathsf{s}) = \mathsf{PreSig}(\tau, m, I_w).$$

\subsection{Pre-verification}

The pre-verification process begins by parsing $S = (\psi(P), \psi(Q))$ and checking the equality of the Weil pairings: $e_{C}(\psi(P), \psi(Q)) = e_{C}(P, Q)^{B}$. Next, using the statement curve $E_w$, extracted from $I_w = (E_w, w(\mathfrak{B}), \pi_w)$, and the commitment curve $E_1$, the proof $\pi_{\psi'}$ is verified by ensuring that:
$1 = \mathsf{NIZK.V}((E_w, E_1), \pi_{\psi'})$,
which confirms that the isogeny $\psi'$ is an isogeny from the statement curve $E_w$ to the curve $E_1$, parallel to the isogeny $\psi: E_0 \to E_\psi$. Subsequently, the challenge isogeny $\varphi = \Phi(E_{\tau}, h): E_{\tau} \to E_2$ is recovered, where $h = \mathsf{H}(j(E_1), m)$. Finally, using the canonical basis $\langle R_1, R_2 \rangle = E_\psi[A]$, it is verified that the representation $\mathcal{R}_{\tilde{\sigma}} = (\tilde{\sigma}(R_1), \tilde{\sigma}(R_2), \deg(\tilde{\sigma}))$ correctly represents an isogeny $\tilde{\sigma}: E_\psi \to E_2$ by computing a higher-dimensional isogeny, as outlined in SQIsignHD. If any of these conditions are not met, the process aborts. The pre-verification algorithm is thus defined as follows: $$0/1 \leftarrow \mathsf{PreVer}(\mathsf{pk}, m, \mathsf{s}, \Tilde{\mathsf{\Sigma}}) = \mathsf{PreVer}\left(E_\tau, m, I_w, (E_1, \pi_{\psi'}, E_{\psi}, S, \mathcal{R}_{\tilde{\sigma}})\right).$$

\begin{figure}[h]
\centering
\begin{tikzcd}[row sep={35,between origins}, column sep={40,between origins}]
 &E_w\arrow[rr, "\psi{'}"]&&E_1\arrow[dddl, bend left, "\sigma"]\\
 E_0 \arrow[ur, "w"]  \arrow[dd, swap, "\tau"] \arrow[rr,"\psi"] && E_{\psi} \arrow[dd, swap, "\tilde{\sigma}"]\arrow[ur, "w'"]  &     \\
 \\
 E_\tau  \arrow[rr, swap, pos=0.4,"\varphi"] &&E_2  & 
\end{tikzcd}
\caption{\textsf{SQIAsignHD} Protocol}
\label{fig:SQIAsignHD Protocol}
\end{figure}

\subsection{Adaptation}

To adapt the pre-signature into a (full) signature, the parallel isogeny $w'$ to the witness isogeny $w$ is first computed using the additional information $S = (\psi(P), \psi(Q))$. This ensures that the resulting second commitment curve, $E_1$, coincides with the codomain of the $w'$, i.e., $w' := [\psi]_{\ast}w: E_\psi \to E_1$, as depicted in Figure \ref{fig:SQIAsignHD Protocol}. 

Next, an efficient representation of the (full) signature isogeny $\sigma := \tilde{\sigma} \circ \hat{w'}: E_1 \to E_2$ is constructed by employing the algorithm $\mathsf{A}$, as described in Definition \ref{def:efficientRepresentation}. The steps are as follows:
\begin{itemize}
    \item[1.] Determine a canonical basis $\langle P_0, Q_0 \rangle := E_1[AC]$.
    \item[2.] Compute $\hat{w'}(P_0)$ and $\hat{w'}(Q_0)$, where $\hat{w'}: E_1 \to E_\psi$ is the dual of $w'$.
    \item[3.] Evaluate $\mathsf{A}(\mathcal{R}_{\tilde{\sigma}}, \hat{w'}(P_0)) =: \sigma(P_0)$ and $\mathsf{A}(\mathcal{R}_{\tilde{\sigma}}, \hat{w'}(Q_0)) =: \sigma(Q_0)$.
    \item[4.] Construct the efficient representation of the isogeny $\sigma: E_1 \to E_2$:
    $$\mathcal{R}_{\sigma} := \big(\sigma(P_0), \sigma(Q_0), \deg(\sigma)\big).$$
\end{itemize}

\noindent The signature is defined as $\mathsf{\Sigma} := (E_1, \mathcal{R}_{\sigma})$. Accordingly, the adaptation algorithm is specified as follows:
\[
\mathsf{\Sigma} := (E_1, \mathcal{R}_{\sigma}) \leftarrow \mathsf{Adapt}(\tilde{\mathsf{\Sigma}}, \mathsf{w}) = \mathsf{Adapt}\big((E_1, \pi_{\psi'}, E_{\psi}, S, \mathcal{R}_{\tilde{\sigma}}), w\big).
\]

\subsection{Extraction} \label{sec:extraction}

In the final phase of our scheme, the goal is to extract the secret witness isogeny $w$ using the publicly known pre-signature $\tilde{\mathsf{\Sigma}}$ and signature $\mathsf{\Sigma}$. This is achieved through two computational approaches: one involves computing the discrete logarithm (of modulus a sufficiently smooth integer), denoted by $\mathsf{\mathcal{A}_{DLP}}$, and the other is an attack for key recovery of an isogeny satisfying $n^2 > 4d$ via the SIDH attack \cite{CDattackSIDH}, denoted by $\mathsf{\mathcal{A}_{SIDH}}$, where $d$ is the degree of the isogeny and $n$ is the order of the given torsion points information. Additionally, we utilize the algorithm $\mathsf{A}$, as defined in Definition \ref{def:efficientRepresentation}. The extraction process proceeds with the following steps:

\begin{itemize}
    \item[1.] Determine a canonical basis $\langle P_1 , Q_1 \rangle = E_1 [N]$ such that $4C < N^2$.
    \item[2.] Set $P' := \mathsf{A}(\mathcal{R}_{\sigma}, P_1)$, $Q' := \mathsf{A}(\mathcal{R}_{\sigma}, Q_1)$, where $P' , Q' \in E_2 [N]$.
    \item[3.] Define $X:= \hat{w'}(P_1)$ and $Y:= \hat{w'}(Q_1)$ as unknowns, for which we seek to determine their values. Then, $X$ and $Y$ can be written as
    $$ X = [a]P_\psi + [b]Q_\psi, \qquad Y = [c]P_\psi + [d]Q_\psi, $$
    for some unknown values $a, b, c, d \in \mathbb{Z}/N\mathbb{Z}$, where $\langle P_\psi , Q_\psi \rangle = E_\psi[N]$.
    \item[4.] From the action of the isogeny $\tilde{\sigma}$ on $X$ and $Y$, we have
    $$ \tilde{\sigma}(X) = \tilde{\sigma}([a]P_\psi + [b]Q_\psi) = [a]\tilde{\sigma}(P_\psi) + [b]\tilde{\sigma}(Q_\psi), $$
    $$ \tilde{\sigma}(Y) = \tilde{\sigma}([c]P_\psi + [d]Q_\psi) = [c]\tilde{\sigma}(P_\psi) + [d]\tilde{\sigma}(Q_\psi), $$
    which gives the following system of equations:
    $$ [a]\tilde{\sigma}(P_\psi) + [b]\tilde{\sigma}(Q_\psi) = P', $$
    $$ [c]\tilde{\sigma}(P_\psi) + [d]\tilde{\sigma}(Q_\psi) = Q', $$
    where $P'$ and $Q'$ were obtained in step 2.
    \item[5.] Set initial values for $a$ and $c$ (we let $a = c = 1$). Using the Discrete Logarithm (DL) algorithm, $\mathsf{\mathcal{A}_{DLP}}$, the values of $b$ and $d$ can be determined. This allows us to determine the action of $\hat{w'}$ on $P_1$ and $Q_1$, i.e., $X$ and $Y$, respectively.
    \item[6.] Apply the SIDH attack, $\mathsf{\mathcal{A}_{SIDH}}$, to find the kernel of the isogeny $\hat{w'}$. Then, compute the dual of $\hat{w'}$, which is the isogeny $w' : E_\psi \to E_1$. 
    \item[7.] Uncover the secret witness $\alpha \in \mathbb{Z}/C\mathbb{Z}$ by expressing the $\ker(w')$ in terms of the already given torsion basis $S = (\psi(P), \psi(Q))$ on $E_{\psi}[C]$, that is, $\ker(w') = \langle \psi(P) + [\alpha] \psi(Q) \rangle$. This is sufficient to recover the witness isogeny via $\ker(w) = \langle P + [\alpha] Q \rangle$, where $P$ and $Q$ are public.
\end{itemize}

\noindent Thus, the extraction algorithm is defined as follows:
$$ w/\perp \leftarrow \mathsf{Ext(\mathsf{\Sigma}, \Tilde{\mathsf{\Sigma}}, s)} = \mathsf{Ext}\Big((E_1, \mathcal{R}_{\sigma}), (E_1, \pi_{\psi'}, E_{\psi}, S, {R}_{\tilde{\sigma}}), I_w \Big).$$

\clearpage
\begin{algorithm}[H]
\caption{$\mathsf{SQIAsignHD}:$ Adaptor Signature $\Xi_{\mathsf{R}_{\frak{A}},\mathsf{\Sigma_{SQIsignHD}}}$}\label{algo1}
\begin{algorithmic}[1]
\State \textbf{Public Parameters.} A prime $p=ABCf -1$, where $A=2^{a}$, $B=\prod_{i=1}^{t}\ell_{i}$, and $C=3^{c}$ are pairwise coprime integers, $f$ is some (small) cofactor, $\ell_{i}$'s are distinct small primes, $A \approx C \approx p^{1/4}$, and $B \approx p^{1/2}$. A supersingular elliptic curve $E_0/\mathbb{F}_{p^{2}}$ with known  $\groupendo(E_0)\cong\mathcal{O}_0 \subset \mathcal{B}_{p,\infty}$, and $|E_0(\mathbb{F}_{p^2})| = (p+1)^2$. An artificial $B$-orientation $\frak{B}=(G_1 , G_2)$ on $E_0$, and a torsion basis $\langle P,Q\rangle=E_{0}[C]$. A secure hash function $\mathsf{H} : \{0, 1\}^* \to [1, \mu(D_\varphi)]$.
\State \textbf{Procedure} $\mathsf{PreSig(sk,m,s)}$ 
\State \indent Parse $I_w$ as  $(E_w,w(\mathfrak{B}),\pi_w)$.
\State \indent Verify that $1=\mathsf{NIZK.V}(E_w,\pi_{w})$.
\State \indent Compute a secret isogeny $\psi: E_0 \to E_\psi$.
\State \indent Compute the image of $P,Q$ under $\psi$, and set $S:=(\psi(P),\psi(Q))$.
\State \indent Compute the push-forward $\psi':=[w]_{\ast}\psi: E_w \to E_1$ via $w(\frak{B})$.
\State \indent Compute the zero-knowledge $\pi_{\psi'}$ showing that $E_1$ is honestly generated.
\State \indent Compute $\varphi:=\Phi(E_\tau ,h): E_\tau \to E_2$, where $h:=\mathsf{H}(j(E_{1}),m)$.
\State \indent Compute $\mathcal{R}_{\tilde{\sigma}}:=(\tilde{\sigma}(R_1),\tilde{\sigma}(R_2),\tilde{q})$ where $\tilde{\sigma}:E_\psi \to E_2$ of degree $\tilde{q}$.
\State \indent $\mathsf{Return}$ $\tilde{\mathsf{\Sigma}}:=(E_1, \pi_{\psi'}, E_\psi, S,\mathcal{R}_{\tilde{\sigma}})$
\State \textbf{Procedure} $\mathsf{PreVer(\mathsf{pk}, m, \mathsf{s}, \Tilde{\mathsf{\Sigma}})}$ 
\State \indent Parse $\tilde{\mathsf{\Sigma}}$ as  $(E_1,\pi_{\psi'}, E_\psi, S,\mathcal{R}_{\tilde{\sigma}})$.
\State \indent Parse $S$ as $(\psi(P),\psi(Q))$.
\State \indent Parse $I_w$ as  $(E_w,w(\mathfrak{B}),\pi_w)$.
\State \indent Check that $e_{C}(\psi(P),\psi(Q)) = e_{C}(P,Q)^{B}$.
\State \indent Verify that $1 = \mathsf{NIZK.V}((E_w, E_1), \pi_{\psi'})$.
\State \indent Recompute $h =\mathsf{H}(j(E_{1}),m)$ and recover $\varphi:=\Phi(E_\tau ,h): E_\tau \to E_2$.
\State \indent Check that $\mathcal{R}_{\tilde{\sigma}}$ correctly represent $\tilde{\sigma}:E_\psi\to E_2$.
\State \indent $\mathsf{Return}$ 0/1.
\State \textbf{Procedure} $\mathsf{Adapt(\tilde{\Sigma},\mathsf{w})}$
\State \indent Compute push-forward $w':=[\psi]_{\ast}w : E_
\psi \to E_1$ via $S$.
\State \indent Determine a canonical basis $\langle P_{0}, Q_{0}\rangle :=E_1 [AC]$.
\State \indent Compute $\sigma(P_0):=\mathsf{A}(\mathcal{R_{\tilde{\sigma}}},\hat{w'}(P_0))$, and $\sigma(Q_0):=\mathsf{A}(\mathcal{R_{\tilde{\sigma}}},\hat{w'}(Q_0))$.
\State \indent Set $\mathcal{R}_{\sigma}:=(\sigma(P_0),\sigma(Q_0),q)$  where $\sigma:E_1 \to E_2$, and $q:=\deg(\sigma)$.
\State \indent $\mathsf{Return}$ $\mathsf{\Sigma}:=(E_1,\mathcal{R}_{\sigma})$
\State \textbf{Procedure} $\mathsf{Ext(\tilde{\Sigma},\Sigma,\mathsf{s})}$
\State \indent Parse $\mathsf{\Sigma}$ as  $(E_1,\mathcal{R}_{\sigma})$.
\State \indent Recover $\hat{w'}:E_1 \to E_\psi$ via $\mathsf{\mathcal{A}_{DLP}}$ and  $\mathsf{\mathcal{A}_{SIDH}}$, and compute $\ker(w')$.
\State \indent Represent $\ker(w')$ in terms of the already given basis $S=(\psi(P), \psi(Q))$.
\State \indent Extract the witness $\alpha \in \mathbb{Z}/C\mathbb{Z}$ for which $\ker(w')=\langle \psi(P) + [\alpha] \psi(Q) \rangle$.
\State \indent $\mathsf{Return}$ $\perp/\,w$
\end{algorithmic}
\label{alg:SQIAsignHD}
\end{algorithm}

\subsection{Parameter Setting}

We follow the parameterization strategy established in SQIsignHD and $\mathsf{bin}\text{SIDH}$, including its hybrid variant $\mathsf{bin}\text{SIDH}^{\mathsf{hyb}}$, to select the underlying prime $p$ in the form $p = ABCf - 1$, where $A = 2^a$, $B = \prod_{i=1}^t \ell_i$, and $C = 3^c$. The parameters are carefully chosen to satisfy $A \approx C \approx p^{1/4}$ and $B \approx p^{1/2}$. Here, the $\ell_i$’s represent distinct small primes greater than 3, and $f$ is a small cofactor.

\textbf{Signature Size.} To achieve 128-bit post-quantum security, the parameter configuration inspired by \textsf{bin}SIDH sets $B$ as the product of the first $134$ primes greater than 3, i.e., $t = 134$. Under this selection, the prime $p$ has an approximate bit length of $|p| \approx 2128$. Moving forward, we examine the detailed structure of the signature. A signature is represented as $\mathsf{\Sigma} = (E_1, \mathcal{R}_{\sigma})$, where $E_1$ denotes a supersingular elliptic curve defined over $\mathbb{F}_{p^2}$, and $\mathcal{R}_{\sigma} = (\sigma(P_0), \sigma(Q_0), \deg(\sigma))$ encodes the image of a canonical torsion basis $(P_0, Q_0)$ of $E_1[AC]$ under the isogeny $\sigma$, along with the degree of the isogeny. The size of the components of the signature is outlined as follows:
\begin{itemize}
\item \textbf{Representation of $E_1$:} The elliptic curve $E_1$ is uniquely determined by its $j$-invariant. For $j(E_1) = a + ib \in \mathbb{F}_{p^{2}}$, storing $j(E_1)$ requires approximately $2\log_2(p)$ bits.
\item \textbf{Isogeny degree:} The degree of the signature isogeny $\sigma$ satisfies $\deg(\sigma) = \deg(\tilde{\sigma}) \cdot \deg(w) \approx p^{3/4}$. Therefore, approximately $\frac{3}{4} \log_2(p)$ bits are required to store the degree.
\item \textbf{Isogeny action on torsion basis:} The images of the torsion points $(P_0, Q_0)$ under $\sigma$ are given by
$$\sigma(P_0) = a_1 Q_1 + b_1 Q_2 \quad \text{and} \quad \sigma(Q_0) = a_2 Q_1 + b_2 Q_2,$$
where $\langle Q_1, Q_2 \rangle = E_2[AC]$ is a canonical basis of the torsion subgroup, and $a_i, b_i \in \mathbb{Z}/AC\mathbb{Z}$ for $i = 1, 2$. Storing these four coefficients requires a total of $4 \log_2(AC) = 2 \log_2(p)$ bits.
\end{itemize}

Summing the contributions, the total signature size amounts to
$$2\log_2(p) + \frac{3}{4}\log_2(p) + 2\log_2(p) = \frac{19}{4}\log_2(p) \text{ bits}.$$
In our setting, this evaluates to approximately $1.26$ KB. To support higher security levels, such as $\lambda \in \{192, 256\}$, one may adopt the parameter scaling strategy suggested by \cite{artificial}, where the number of small primes used in the construction of $B$ is increased proportionally. In particular, it is reasonable to set $t = \lambda$, thereby ensuring that the calculation is made while maintaining a balance among the parameters similar to the 128-bit configuration.

\begin{remark}
    The pre-signature $\tilde{\mathsf{\Sigma}}$ incorporates a zero-knowledge proof for the commitment isogeny. Although the pre-signature is inherently ephemeral, the size of the zero-knowledge proof remains an important consideration, primarily influenced by the underlying isogeny structure. For artificially oriented curves, the construction adapts the zero-knowledge proof for masked public keys from \cite{oprf_zk_basso} to accommodate independently scaled points. While this adaptation preserves the desired security properties, its efficiency—particularly in terms of proof size and computational cost—remains an area for improvement. Investigating more compact encoding methods or alternative proof techniques may enhance the overall practicality and scalability of the scheme.
\end{remark}

\section{Security Proof}\label{sec4}
In this section, we analyze and formally prove the security of the proposed adaptor signature scheme, denoted as $\Xi_\mathsf{R_{\mathfrak{A}},\Sigma_{SQIsignHD}}$, as introduced in Algorithm \ref{alg:SQIAsignHD}. We demonstrate that $\Xi_\mathsf{R_{\mathfrak{A}},\Sigma_{SQIsignHD}}$ satisfies the properties of pre-signature correctness, pre-signature adaptability, \textsf{aEUF-CMA}, and witness extractability. Verifying these properties is sufficient to prove Theorem \ref{securitytheorem}.

\begin{lemma}\label{lemma:securityLemma1}
    The adaptor signature $\Xi_\mathsf{R_{\mathfrak{A}},\Sigma_{SQIsignHD}}$, as presented in Algorithm \ref{alg:SQIAsignHD}, is pre-signature correct. 
    \begin{proof}
        First, let $\mathsf{(w,s)}:= (w, I_w = (E_w, w(\mathfrak{B}), \pi_w))\xleftarrow{\$} \mathsf{GenR}(1^{\lambda})$ represent a fixed witness/statement pair for the defined hard relation $\mathsf{R_{\mathfrak{A}}}$. Here, $w$ denotes an isogeny from $E_0$ to the target elliptic curve $E_w$, $w(\mathfrak{B})$ represents the image of the $B$-orientation $\mathfrak{B}$ under the witness isogeny $w$, and $\pi_w$ is a zero-knowledge proof for the pair $(w, (E_w,(\mathfrak{B})))$. Additionally, let $(\mathsf{sk}, \mathsf{pk}):= (\tau, E_\tau)\xleftarrow{\$} \mathsf{KeyGen}(1^{\lambda})$ be a fixed secret/public key pair.

        \noindent Assume that, for a message $m \in \{0,1\}^{\ast}$, the pre-signature $\tilde{\mathsf{\Sigma}} = (E_1, \pi_{\psi'}, E_{\psi}, S, \mathcal{R}_{\tilde{\sigma}})$ is generated via the $\mathsf{PreSig}$ algorithm, i.e., $\tilde{\mathsf{\Sigma}} \leftarrow \mathsf{PreSig}(\tau, m, I_w)$. In this case, the verification algorithm yields $1 \leftarrow \mathsf{PreVer}(E_\tau, m, I_w, \tilde{\mathsf{\Sigma}})$. This holds because: (1) $\mathcal{R}_{\tilde{\sigma}}$ is a correct efficient representation of an isogeny $\tilde{\sigma}$ from $E_\psi$ to $E_2$, constructed using knowledge of $\text{End}(E_0)$ and the isogenies $\tau$, $\psi$, and $\varphi$; and (2) the isogeny $\varphi:E_\tau\to E_2$ depends on the message $m$ and the $j$-invariant of the (second) commitment curve $E_1$. The curve $E_1$ is obtained by pushing forward the commitment isogeny $\psi$ through the witness isogeny $w$, i.e., $[w]_{\ast}\psi : E_w \to E_1$. By the correctness of \textsf{NIZK}, we have $1 = \mathsf{NIZK.V}(E_1, \pi_{\psi'})$. Furthermore, for $S=(\psi(P), \psi(Q))$, the equality of the Weil pairings $e_{C}(\psi(P), \psi(Q)) = e_{C}(P, Q)^{B}$ holds.

        \noindent Next, consider the (full) signature $\mathsf{\Sigma} = (E_{1}, \mathcal{R}_{\sigma})$ produced by the adaptation algorithm, i.e., $\mathsf{\Sigma} \leftarrow \mathsf{Adapt}(\tilde{\mathsf{\Sigma}}, w)$. The verification algorithm $\mathsf{Ver}$ of $\mathsf{\Sigma_{SQIsignHD}}$ returns $1 \leftarrow \mathsf{Ver}(E_{\tau}, m, \mathsf{Adapt}(\tilde{\mathsf{\Sigma}}, w))$, since $\mathcal{R}_{\sigma}$ is an efficient representation of the signature isogeny $$\sigma:= \tilde{\sigma} \circ \widehat{[\psi]_{\ast} w} = \tilde{\sigma} \circ \hat{w'} : E_1 \rightarrow E_{2},$$ 
        where $E_1$ is derived by pushing forward the witness isogeny $w$ through $\psi$ using $S$, and $E_2$ is the codomain of isogeny $\varphi$ induced by $j(E_1)$ and the message $m$.
        
        \noindent Using the pre-signature $\tilde{\mathsf{\Sigma}} = (E_1, \pi_{\psi'}, E_{\psi}, S, \mathcal{R}_{\tilde{\sigma}})$ and the signature $\mathsf{\Sigma} = (E_{1}, \mathcal{R}_{\sigma})$, we can exploit the discrete logarithm algorithm $\mathsf{\mathcal{A}_{DLP}}$ and the SIDH attack $\mathsf{\mathcal{A}_{SIDH}}$ to extract the isogeny $w' : E_\psi \rightarrow E_{1}=E_\psi/\langle \psi(P)+[\alpha]\psi(Q)\rangle$, as described in Section \ref{sec:extraction}. The secret value $\alpha$ then suffices to construct the witness isogeny $w : E_0 \rightarrow E_{w}=E_{0}/\langle P+[\alpha]Q\rangle$. Consequently, $w \leftarrow \mathsf{Ext}\big(\mathsf{\Sigma}, \tilde{\mathsf{\Sigma}}, I_w\big)$ can be successfully executed to recover the secret witness isogeny $w$. Therefore, the adaptor signature $\Xi_{\mathsf{R_{\mathfrak{A}}, \Sigma_{SQIsignHD}}}$ satisfies the pre-signature correctness property.
        \end{proof}
\end{lemma}

\begin{lemma}\label{lemma:securityLemma2}
    The adaptor signature $\Xi_\mathsf{R_{\mathfrak{A}},\Sigma_{SQIsignHD}}$, as depicted in Algorithm \ref{alg:SQIAsignHD}, is pre-signature adaptable. 
    \begin{proof}
        Let us define a fixed witness/statement pair $\mathsf{(w, s)} := (w, I_w) \in \mathsf{R_{\mathfrak{A}}}$, a fixed public key $\mathsf{pk} = E_\tau$, a pre-signature $\tilde{\mathsf{\Sigma}}$, and a message $m \in \{0,1\}^{\ast}$, as in Lemma \ref{lemma:securityLemma1}.

        \noindent We aim to prove that any verifiably valid (though not necessarily honestly generated) pre-signature $\tilde{\mathsf{\Sigma}} = (E_1, \pi_{\psi’}, E_{\psi}, S, \mathcal{R}_{\tilde{\sigma}})$ that passes the $\mathsf{PreVer}$ procedure can be adapted into a valid (full) signature $\mathsf{\Sigma}$. \\
        Assuming $\mathsf{PreVer}(E_\tau, m, I_w, \tilde{\mathsf{\Sigma}}) = 1$, it follows from the pre-verification procedure that $\mathsf{NIZK.V}(E_1, \pi_{\psi’}) = 1$, the equality $e_{C}(\psi(P), \psi(Q)) = e_{C}(P, Q)^{B}$ of Weil pairings holds, and $\mathcal{R}_{\tilde{\sigma}}$ represents an isogeny from $E_\psi$ to $E_2$, where $E_2$ is the target curve of $\varphi$, derived from (the hash of) the message $m$ and the $j$-invariant of commitment curve $E_1$. By the correctness property established in Lemma \ref{lemma:securityLemma1}, and given the presence of the witness $w$ corresponding to the statement $I_w$, the adaptation algorithm $\mathsf{Adapt}$ necessarily produces a full signature $\mathsf{\Sigma}$ by first computing the push-forward $w'=[\psi]_{\ast}w:E_{\psi}\to E_{1}$ using $S=(\psi(P),\psi(Q))$, and then computing the composition $\sigma=\tilde{\sigma}\circ\hat{w}':E_{1}\to E_2$ to produce the efficient representation $\mathcal{R}_{\sigma}$. Consequently, the verification algorithm $\mathsf{Ver}$ of $\mathsf{\Sigma_{SQIsignHD}}$ necessarily accepts the signature $\mathsf{\Sigma} = (E_{1}, \mathcal{R}_{\sigma})$, i.e., $1 \leftarrow \mathsf{Ver}(E_{\tau}, m, \mathsf{Adapt}(\tilde{\mathsf{\Sigma}}, w))$.
    \end{proof}
\end{lemma}

\begin{lemma}\label{lemma:securityLemma3}
    Assuming that the SQIsignHD signature scheme $\mathsf{\Sigma_{SQIsignHD}}$ is $\mathsf{SUF}$-$\mathsf{CMA}$-secure, that $\mathsf{R_{\mathfrak{A}}}$ is a hard relation, and that Problem \ref{problem:SSIP-A} and Problem \ref{problem:SSIP-B} are computationally hard, then the $\mathsf{SQIAsignHD}$ adaptor signature scheme $\Xi_\mathsf{R_{\mathfrak{A}},\Sigma_{SQIsignHD}}$, as given in Algorithm \ref{alg:SQIAsignHD}, is $\mathsf{aEUF\text{-}CMA}$-secure.
    \begin{proof}
    We begin our proof by reducing the unforgeability of the $\mathsf{SQIAsignHD}$ adaptor signature scheme to the strong unforgeability of the SQIsignHD signature scheme. Specifically, we consider an adversary $\mathcal{A}$ who plays a series of games, starting with the \textsf{aSigForge} game as defined in Definition \ref{def:aEUF-CMA}. We then construct a simulator $\mathcal{S}$ who plays the strong unforgeability experiment \textsf{StrongSigForge}, as defined in Definition \ref{def:StrongSigForge} for the SQIsignHD signature scheme. The simulator $\mathcal{S}$ leverages $\mathcal{A}$’s forgery in \textsf{aSigForge} to win its own experiment. In this setting, $\mathcal{S}$ has access to both the signing oracle $\mathsf{Sig^{SQIsignHD}}$ and the random oracle $\mathcal{H^{\mathsf{SQIsignHD}}}$, which it uses to simulate oracle queries for $\mathcal{A}$: specifically, the random oracle $\mathcal{H}$, the signing queries $\mathcal{O}_S$, and the pre-signing queries $\mathcal{O}_{pS}$.

    The primary challenges in simulating oracles arise when handling $\mathcal{O}_{pS}$ queries. Since $\mathcal{S}$ can only obtain full signatures from its signing oracle, it requires a method to transform these full signatures into pre-signatures suitable for $\mathcal{A}$. This transformation process presents two main difficulties: (1) $\mathcal{S}$ must learn the witness $w$ corresponding to the statement $I_w$ for which the pre-signature is to be generated, and (2) $\mathcal{S}$ must simulate the zero-knowledge proof $\pi_{\psi’}$ associated with a secret parallel isogeny $\psi'$ of the commitment isogeny $\psi$, ensuring consistency in the randomness within the pre-signature.

    More specifically, upon receiving a $\mathcal{O}_{pS}$ query from $\mathcal{A}$, which includes a message $m$ and an instance $I_w = (E_w, w(\mathfrak{B}), \pi_w)$, the simulator $\mathcal{S}$ queries its signing oracle $\mathsf{Sig^{SQIsignHD}}$  to obtain a full signature on $m$. Furthermore, the simulator must learn a witness $w$ such that $(w, I_w)\in \mathsf{R_{\mathfrak{A}}}$ in order to convert the full signature into a pre-signature for $\mathcal{A}$. To this end, we utilize the extractability property of the zero-knowledge proof $\pi_w$, which allows us to extract $w$ and, in turn, transform the full signature into a valid pre-signature. Additionally, since a valid pre-signature includes a zero-knowledge proof $\pi_{\psi’}$, the simulator must simulate this proof without knowledge of the corresponding secret. To achieve this, we rely on the zero-knowledge property, which enables the simulation of a proof for a statement without requiring access to the associated witness.\\
    
    \noindent $\mathsf{Game_0}$. This game corresponds to the $\mathsf{aSigForge}$ experiment, as per Definition \ref{def:aEUF-CMA}, where the adversary $\mathcal{A}$ has access to a random oracle $\mathcal{H}$ in the random oracle model, as well as many previously produced valid pre-signatures and signatures through the pre-signing oracle $\mathcal{O}_{pS}$ and the signing oracle $\mathcal{O}_S$, respectively, for messages of its choice, except for a message $m$. The adversary then attempts to forge a verifiable signature $\mathsf{\Sigma}^\ast$ on $m$. Since we are working within the random oracle model, we explicitly write the random oracle code $\mathcal{H}$ via $\mathcal{H}^{\mathsf{SQIsignHD}}$. Thus, it follows that 
    $$\text{Pr}[\mathsf{Game_0}=1]=\text{Pr}[\mathsf{aSigForge}_{{\mathcal{A},\Xi_{\mathsf{R_{\mathfrak{A}},\mathsf{\Sigma_{SQIsignHD}}}}}}(\lambda)=1].$$

    \begin{pchstack}[boxed, center, space=1em]\label{fig:aEUFgame0}
    \begin{pcvstack}[space=.3em]
    \procedure[linenumbering]{\pcgame [0]}{%
    \mathcal{Q}:=\emptyset \\
    H:=[\perp]\\
    (\tau,E_\tau)\leftarrow \mathsf{KeyGen}(\secparam)\\
    m\leftarrow \adv^{\mathcal{H},\mathcal{O}_{S},\mathcal{O}_{pS}}(E_\tau)\\
    (w,I_w)\leftarrow \mathsf{GenR(1^{\lambda})}\\
    \tilde{\mathsf{\Sigma}}\leftarrow\mathsf{PreSig}(\tau,m,I_w)\\
    \mathsf{\Sigma^\ast}\leftarrow\adv^{\mathcal{H},\mathcal{O}_{S},\mathcal{O}_{pS}}(\tilde{\mathsf{\Sigma}},I_w)\\
    b:=\mathsf{Ver}(E_\tau ,m,\mathsf{\Sigma^\ast})\\
    \pcreturn m\not\in\mathcal{Q}\pmb{\land} b
    }
    \end{pcvstack}

    \begin{pcvstack}[space=0.3em]
    
    \procedure[linenumbering]{$\mathcal{H} (x)$}{%
    \pcif H[x]=\perp\\
    \pcind H[x]\leftarrow\mathcal{H}^{\mathsf{SQIsignHD}}(x)\\
    \pcreturn H[x]
    }
    \procedure[linenumbering]{$\mathcal{O}_{pS} (m,I_w)$}{%
    \tilde{\mathsf{\Sigma}}\leftarrow\mathsf{PreSig}(\tau,m,I_w)\\
    \mathcal{Q}:=\mathcal{Q}\cup\{m\}\\
    \pcreturn \tilde{\mathsf{\Sigma}}
    }
    \procedure[linenumbering]{$\mathcal{O}_S \big(m)$}{%
    \mathsf{\Sigma}\leftarrow\mathsf{Sig}(\tau, m) \\
    \mathcal{Q}:=\mathcal{Q}\cup\{m\} \\
    \pcreturn \mathsf{\Sigma}
    }
    \end{pcvstack}
\end{pchstack}

\begin{pchstack}[boxed, center, space=1em]\label{fig:aEUFgame1}
    \begin{pcvstack}[space=.3em]
    \procedure[linenumbering]{\pcgame [1]}{%
    \mathcal{Q}:=\emptyset \\
    H:=[\perp]\\
    (\tau,E_\tau)\leftarrow \mathsf{KeyGen}(\secparam)\\
    m^\ast\leftarrow \adv^{\mathcal{H},\mathcal{O}_{S},\mathcal{O}_{pS}}(E_\tau)\\
    (w,I_w)\leftarrow \mathsf{GenR(1^{\lambda})}\\
    \tilde{\mathsf{\Sigma}}\leftarrow\mathsf{PreSig}(\tau,m^\ast,I_w)\\
    \mathsf{\Sigma^\ast}\leftarrow\adv^{\mathcal{H},\mathcal{O}_{S},\mathcal{O}_{pS}}(\tilde{\mathsf{\Sigma}},I_w)\\
    \gamechange{$\pcif \mathsf{Adapt(\tilde{\Sigma}},w)=\mathsf{\Sigma}^\ast$}\\
    \pcind \gamechange{\pcabort}\\
    b:=\mathsf{Ver}(E_\tau ,m^\ast,\mathsf{\Sigma^\ast})\\
    \pcreturn m^\ast\not\in\mathcal{Q}\pmb{\land} b
    }
    \end{pcvstack}

    \begin{pcvstack}[space=0.3em]
    
    \procedure[linenumbering]{$\mathcal{H} (x)$}{%
    \pcif H[x]=\perp\\
    \pcind H[x]\leftarrow\mathcal{H}^{\mathsf{SQIsignHD}}(x)\\
    \pcreturn H[x]
    }
    \procedure[linenumbering]{$\mathcal{O}_{pS} (m,I_w)$}{%
    \tilde{\mathsf{\Sigma}}\leftarrow\mathsf{PreSig}(\tau,m,I_w)\\
    \mathcal{Q}:=\mathcal{Q}\cup\{m\}\\
    \pcreturn \tilde{\mathsf{\Sigma}}
    }
    \procedure[linenumbering]{$\mathcal{O}_S \big(m)$}{%
    \mathsf{\Sigma}\leftarrow\mathsf{Sig}(\tau, m) \\
    \mathcal{Q}:=\mathcal{Q}\cup\{m\} \\
    \pcreturn \mathsf{\Sigma}
    }
    \end{pcvstack}
\end{pchstack}

\noindent $\mathsf{Game_1}$. This game is analogous to $\mathsf{Game_0}$, with the only difference being that if the valid signature $\mathsf{\Sigma}^{\ast}$, forged by the adversary $\mathcal{A}$, matches the result of adapting the pre-signature into a signature using the corresponding witness $w$, then the game aborts.

\begin{claim}
    If $\mathsf{Bad_1}$ is the event that $\mathsf{Game_1}$ aborts, then we claim that for a negligible function $\mathsf{negl}$ in $\lambda$, \text{Pr}$[\mathsf{Bad_1}]\leq\mathsf{negl}(\lambda)$.
    \begin{proof}
        We prove this claim by reducing it to the hardness of the relation $\mathsf{R_{\mathfrak{A}}}$. To do this, we construct a simulator $\mathcal{S}$ that breaks the hardness of $\mathsf{R_{\mathfrak{A}}}$ under the assumption that it has access to an adversary $\mathcal{A}$ that causes $\mathsf{Game_1}$ to abort with non-negligible probability. The simulator receives a challenge $\mathsf{s}^\ast:=I_{w^\ast}^{\ast}$, upon which it generates a secret/public key pair $(\tau, E_\tau)\leftarrow \mathsf{KeyGen(1^\lambda)}$ to simulate $\mathcal{A}$’s queries to the oracles $\mathcal{H}$, $\mathcal{O}_{pS}$ and $\mathcal{O}_S$. The simulation of the oracles proceeds as described in $\mathsf{Game_1}$. 
        
        Upon receiving the challenge message $m^\ast$ from $\mathcal{A}$, $\mathcal{S}$ computes a pre-signature $\tilde{\mathsf{\Sigma}}\leftarrow \mathsf{PreSig}(\tau, m^\ast,  I_{w^\ast}^{\ast})$ and returns the pair $(\tilde{\mathsf{\Sigma}}, I_{w^\ast}^{\ast})$ to the adversary, who forges a signature using the returned pair. Assuming that $\mathsf{Bad_1}$ occurred (i.e., $\mathsf{Adapt(\tilde{\Sigma}},w) = \mathsf{\Sigma}^\ast)$. Since the $\Xi_\mathsf{R_{\mathfrak{A}},\Sigma_{SQIsignHD}}$ is pre-signature correct by Lemma \ref{lemma:securityLemma1}, the simulator can extract $w^\ast$ via $\mathsf{Ext(\Sigma^\ast, \tilde{\Sigma}}, I_{w^\ast}^{\ast})$ to obtain a valid witness/statement pair such that $(w^\ast, I_{w^\ast}^{\ast})\in\mathsf{R_{\mathfrak{A}}}$. In this way, $\mathcal{S}$ breaks the security of the relation $\mathsf{R_{\mathfrak{A}}}$.
        
        We note that the view of $\mathcal{A}$ is indistinguishable from its view in $\mathsf{Game_1}$, since the challenge $I_{w^\ast}^{\ast}$ is an instance of the hard relation $\mathsf{R_{\mathfrak{A}}}$ and follows the same distribution as the public output of $\mathsf{GenR}$. Therefore, the probability that $\mathcal{S}$ breaks the hardness of $\mathsf{R_{\mathfrak{A}}}$ is equal to the probability that the event $\mathsf{Bad_1}$ occurring, which is non-negligible by assumption. This contradicts the hardness of $\mathsf{R_{\mathfrak{A}}}$. 
\end{proof}
\end{claim} 

Since $\mathsf{Game_1}$ and $\mathsf{Game_0}$ are equivalent except when the event $\mathsf{Bad_1}$ occurs, it follows that
$$\text{Pr}[\mathsf{Game_1} = 1] \leq \text{Pr}[\mathsf{Game_0} = 1] + \mathsf{negl(\lambda)}.$$

\noindent $\mathsf{Game_2}$. This game is similar to the previous game, with the only difference being a modification in the pre-signing oracle $\mathcal{O}_{pS}$. 
Specifically, in this game, we apply the extractor algorithm $\mathcal{E}$, taking the statement $(E_w, w(\mathfrak{B}))$, the proof $\pi_w$, and the list of random oracle queries $H$ as input to extract a witness $w$. The game aborts if $(w, (E_w, w(\mathfrak{B}),\pi_w))\not\in\mathsf{R_{\mathfrak{A}}}$.
\begin{claim}
If $\mathsf{Bad_2}$ is the event that \noindent $\mathsf{Game_2}$ aborts during an $\mathcal{O}_{pS}$ execution, then it
holds that $Pr\mathsf{[Bad_2]} \leq \mathsf{negl(\lambda)}$, for a negligible function $\mathsf{negl}$ in $\lambda$.  
\begin{proof}
By the online extractor property of the \textsf{NIZK}, for a witness $w$ extracted from a proof $\pi_w$ of the statement $(E_w, w(\mathfrak{B}))$ such that $\mathsf{NIZK.V}(E_w, w(\mathfrak{B}), \pi_w) = 1$, it follows that $(w, I_w) \in \mathsf{R_{\mathfrak{A}}}$, except with negligible probability in the security parameter $\lambda$.
\end{proof}
\end{claim}

Therefore, since games $\mathsf{Game_2}$ and $\mathsf{Game_1}$ are equivalent except in case event $\mathsf{Bad_2}$ happens, it follows that
$$Pr[\mathsf{Game_2} = 1] \leq Pr[\mathsf{Game_1} = 1] + \mathsf{negl}(\lambda).$$

\begin{pchstack}[boxed, center, space=1em]
\begin{pcvstack}[space=.3em]
    \procedure[linenumbering]{\pcgame [2]}{%
    \mathcal{Q}:=\emptyset \\
    H:=[\perp]\\
    (\tau,E_\tau)\leftarrow \mathsf{KeyGen}(\secparam)\\
    m^\ast\leftarrow \adv^{\mathcal{H},\mathcal{O}_{S},\mathcal{O}_{pS}}(E_\tau)\\
    (w,I_w)\leftarrow \mathsf{GenR(1^{\lambda})}\\
    \tilde{\mathsf{\Sigma}}\leftarrow\mathsf{PreSig}(\tau,m^\ast,I_w)\\
    \mathsf{\Sigma^\ast}\leftarrow\adv^{\mathcal{H},\mathcal{O}_{S},\mathcal{O}_{pS}}(\tilde{\mathsf{\Sigma}},I_w)\\
    \pcif \mathsf{Adapt(\tilde{\Sigma}},w)=\mathsf{\Sigma}^\ast\\
    \pcind \pcabort\\
    b:=\mathsf{Ver}(E_\tau ,m^\ast,\mathsf{\Sigma^\ast})\\
    \pcreturn m^\ast\not\in\mathcal{Q}\pmb{\land} b
    }
    \end{pcvstack}

    \begin{pcvstack}[space=.3em]
    \procedure[linenumbering]{$\mathcal{H} (x)$}{%
    \pcif H[x]=\perp\\
    \pcind H[x]\leftarrow\mathcal{H}^{\mathsf{SQIsignHD}}(x)\\
    \pcreturn H[x]
    }
    \procedure[linenumbering]{$\mathcal{O}_{pS} (m,I_w)$}{%
    \gamechange{Parse $I_w$ as $(E_w,w(\frak{B}),\pi_w)$}\\
    \gamechange{$w := \mathcal{E}(E_w,w(\frak{B}),\pi_w, H)$}\\
    \gamechange{$\pcif (w,I_w)\not\in \mathsf{R_{\mathfrak{A}}}$}\\
    \pcind\gamechange{$\pcabort$}\\
    \tilde{\mathsf{\Sigma}}\leftarrow\mathsf{PreSig}\big(\tau,m,I_w\big)\\
    \mathcal{Q}:=\mathcal{Q}\cup\{m\}\\
    \pcreturn \tilde{\mathsf{\Sigma}}
    }
    \procedure[linenumbering]{$\mathcal{O}_S (m)$}{%
    \mathsf{\Sigma}\leftarrow\mathsf{Sig}(\tau,m)\\
    \mathcal{Q}:=\mathcal{Q}\cup\{m\} \\
    \pcreturn \mathsf{\Sigma}
    }
    \end{pcvstack}
\end{pchstack}

\begin{pchstack}[boxed, center, space=.3em]
    \begin{pcvstack}[space=.3em]
    \procedure[linenumbering]{\pcgame [3]}{%
    \mathcal{Q}:=\emptyset \\
    H:=[\perp]\\
    (\tau,E_\tau)\leftarrow \mathsf{KeyGen}(\secparam)\\
    m^\ast\leftarrow \adv^{\mathcal{H},\mathcal{O}_{S},\mathcal{O}_{pS}}(E_\tau)\\
    (w,I_w)\leftarrow \mathsf{GenR(1^{\lambda})}\\
    \tilde{\mathsf{\Sigma}}\leftarrow\mathsf{PreSig}(\tau,m^\ast,I_w)\\
    \mathsf{\Sigma^\ast}\leftarrow\adv^{\mathcal{H},\mathcal{O}_{S},\mathcal{O}_{pS}}(\tilde{\mathsf{\Sigma}},I_w)\\
    \pcif \mathsf{Adapt(\tilde{\Sigma}},w)=\mathsf{\Sigma}^\ast\\
    \pcind \pcabort\\
    b:=\mathsf{Ver}(E_\tau ,m^\ast,\mathsf{\Sigma^\ast})\\
    \pcreturn m^\ast\not\in\mathcal{Q}\pmb{\land} b
    }
    
    \procedure[linenumbering]{$\mathcal{O}_S (m)$}{%
    \mathsf{\Sigma}\leftarrow\mathsf{Sig}(\tau,m)\\
    \mathcal{Q}:=\mathcal{Q}\cup\{m\} \\
    \pcreturn \mathsf{\Sigma}
    }
    \end{pcvstack}

    \begin{pcvstack}[space=.3em]
    
    \procedure[linenumbering]{$\mathcal{H} (x)$}{%
    \pcif H[x]=\perp\\
    \pcind H[x]\leftarrow\mathcal{H}^{\mathsf{SQIsignHD}}(x)\\
    \pcreturn H[x]
    }
    \procedure[linenumbering]{$\mathcal{O}_{pS} (m,I_w)$}{%
    \text{Parse } I_w \text{ as } (E_w,w(\frak{B}),\pi_w)\\
    w := \mathcal{E}(E_w,w(\frak{B}), \pi_w , H)\\
    \pcif (w, I_w)\not\in \mathsf{R_{\mathfrak{A}}}\\
    \pcind \pcabort\\
\gamechange{$\mathsf{\Sigma}\leftarrow\mathsf{Sig}(\tau,m)$}\\
\gamechange{$\text{Parse }\mathsf{\Sigma} \text{ as } (E_{1},\mathcal{R}_{\sigma})$}\\
\gamechange{\text{Extract} $(E^{\ast}_{\psi},\mathcal{R}^{\ast}_{\tilde{\sigma}}) \text{ by } \mathcal{A}_{\mathsf{SIDH}} \text{ s.t. }$}\\
\pcind \gamechange{$\sigma = \tilde{\sigma}\circ \hat{w}^{\ast}$ with $\deg(\hat{w}^{\ast})=C$}\\
\gamechange{$\text{Extract }\alpha \text{ from }$}\\
\pcind \gamechange{$\ker(w)=\langle P+[\alpha]Q\rangle$}\\
\gamechange{Find $\langle P^{\ast},Q^{\ast}\rangle=E^{\ast}_{\psi}[C]$ for which}\\
\pcind\gamechange{$\ker(w^{\ast})=\langle P^{\ast}+[\alpha]Q^{\ast}\rangle$}\\
%\pcind\gamechange{$e_{C}(P,Q)=e_{C}(P^{\ast},Q^{\ast})^{B}$}\\
\gamechange{Set $S^{\ast}:=(P^{\ast},Q^{\ast})$}\\
\gamechange{$\pi^{\ast}_{\psi'}\leftarrow \mathcal{S}((E_w, E_1),1)$}\\
\mathcal{Q}:=\mathcal{Q}\cup\{m\}\\
\gamechange{$\pcreturn \tilde{\mathsf{\Sigma}}:=(E_1, \pi^{\ast}_{\psi'}, E^{\ast}_{\psi},S^{\ast},\mathcal{R}^{\ast}_{\tilde{\sigma}})$}
    }
    \end{pcvstack}
\end{pchstack} 

\noindent $\mathsf{Game_3}$. This game extends the modifications to the pre-signing oracle $\mathcal{O}_{pS}$ introduced in the previous game. Specifically, it begins by generating a valid full signature $\mathsf{\Sigma}=(E_1,\mathcal{R}_{\sigma})$ through the execution of the $\mathsf{Sig}$ algorithm. Using the $\mathcal{A}_{\mathsf{SIDH}}$ algorithm, the isogeny $\sigma$, represented by $\mathcal{R}_{\sigma}$, is  decomposed into $\sigma=\tilde{\sigma}\circ \hat{w}^{\ast}$, where $\hat{w}^{\ast}$ is a $C$-isogeny from $E_1$ to a curve $E^{\ast}_{\psi}$, and $\tilde{\sigma}$ is an isogeny from $E^{\ast}_{\psi}$ to $E_2$. Subsequently, the efficient representation $\mathcal{R}_{\tilde{\sigma}}$ corresponding to the isogeny $\tilde{\sigma}$ is computed.

Now, in order to construct a $C$-torsion basis $\langle P^{\ast}, Q^{\ast} \rangle$ for $E_{\psi}^{\ast}[C]$ for which $\ker(w^{\ast}) = \langle P^{\ast} + [\alpha]Q^{\ast} \rangle$, where $\alpha$ is the secret obtained from the extracted witness $w: E_0 \to E_0 / \langle P + [\alpha]Q \rangle$, and $w^{\ast}$ is the dual of $\hat{w}^{\ast}$, let $\ker(w^{\ast}) = R$. First, we select a point $R’$ that is linearly independent of $R$ to form a $C$-torsion basis for $E_{\psi}^{\ast}$, i.e., $\langle R, R’ \rangle = E_{\psi}^{\ast}[C]$. Now, we seek values $x_1, y_1, x_2$, and $y_2$ such that
$$(x_1R + y_1R’) + \alpha(x_2R + y_2R’) = R.$$
This condition implies:
$$x_1 + \alpha x_2 = 1 \quad \text{and} \quad y_1 + \alpha y_2 = 0.$$

Finding a single solution for $(x_1, x_2), (y_1, y_2) \in (\mathbb{Z}/C\mathbb{Z}) \times (\mathbb{Z}/C\mathbb{Z})$, where $(x_i, y_i) \neq (0, 0)$ for $i = 1, 2$, is sufficient to determine the pair $S^{\ast}=(P^{\ast},Q^{\ast})$ by setting:
$$P^{\ast} := x_1R + y_1R’ \quad \text{and} \quad Q^{\ast} := x_2R + y_2R’.$$

Finally, before forming the pre-signature, the $\mathcal{S}$ simulates a proof $\pi^{\ast}_{\psi'}$ for the statement $E_1$ without any knowledge of the corresponding secret isogeny $\psi'$. The pre-signature is then defined as $\tilde{\mathsf{\Sigma}}:=(E_1, \pi^{\ast}_{\psi'}, E^{\ast}_{\psi},S^{\ast},\mathcal{R}^{\ast}_{\tilde{\sigma}})$. We see that this game is indistinguishable from the previous one, and it follows that
$$Pr[\mathsf{Game_3} = 1] \leq Pr[\mathsf{Game_2} = 1] + \mathsf{negl}(\lambda).$$

    \begin{pchstack}[boxed, center, space=0.3em]\label{fig:aEUFgame4}
    \footnotesize{
    \begin{pcvstack}[space=.3em]
    \procedure[linenumbering]{\pcgame [4]}{%
    \mathcal{Q}:=\emptyset \\
    H:=[\perp]\\
    (\tau,E_\tau)\leftarrow \mathsf{KeyGen}(\secparam)\\
    m^\ast\leftarrow \adv^{\mathcal{H},\mathcal{O}_{S},\mathcal{O}_{pS}}(E_\tau)\\
    (w,I_{w})\leftarrow \mathsf{GenR(1^{\lambda})}\\
\gamechange{$\mathsf{\Sigma}\leftarrow\mathsf{Sig}(\tau,m^{\ast})$}\\
\gamechange{$\text{Parse }\mathsf{\Sigma} \text{ as } (E_{1},\mathcal{R}_{\sigma})$}\\
\gamechange{\text{Extract} $(E^{\ast}_{\psi},\mathcal{R}^{\ast}_{\tilde{\sigma}}) \text{ by } \mathcal{A}_{\mathsf{SIDH}} \text{ s.t. }$}\\
\pcind \gamechange{$\sigma = \tilde{\sigma}\circ \hat{w}^{\ast}$ with $\deg(\hat{w}^{\ast})=C$}\\
\gamechange{$\text{Extract }\alpha \text{ from }$}\\
\pcind \gamechange{$\ker(w)=\langle P+[\alpha]Q\rangle$}\\
\gamechange{Find $\langle P^{\ast},Q^{\ast}\rangle=E^{\ast}_{\psi}[C]$ for which}\\
\pcind\gamechange{$\ker(w^{\ast})=\langle P^{\ast}+[\alpha]Q^{\ast}\rangle$}\\
%\pcind\gamechange{$e_{C}(P,Q)=e_{C}(P^{\ast},Q^{\ast})^{B}$}\\
\gamechange{Set $S^{\ast}:=(P^{\ast},Q^{\ast})$}\\
\gamechange{$\pi^{\ast}_{\psi'}\leftarrow \mathcal{S}((E_w, E_1),1)$}\\
\gamechange{$\tilde{\mathsf{\Sigma}}:=(E_1, \pi^{\ast}_{\psi'}, E^{\ast}_{\psi},S^{\ast},\mathcal{R}^{\ast}_{\tilde{\sigma}})$}\\
\gamechange{$\mathsf{\Sigma^\ast}\leftarrow\adv^{\mathcal{H},\mathcal{O}_{S},\mathcal{O}_{pS}}(\tilde{\mathsf{\Sigma}},I_w)$}\\
    \pcif \mathsf{Adapt(\tilde{\Sigma}},w)=\mathsf{\Sigma}^\ast\\
    \pcind \pcabort\\
    b:=\mathsf{Ver}(E_\tau,m^\ast,\mathsf{\Sigma^\ast})\\
    \pcreturn m^\ast\not\in\mathcal{Q}\land b
    }
    \end{pcvstack}

    \begin{pcvstack}[space=.3em]
    
    \procedure[linenumbering]{$\mathcal{H} (x)$}{%
    \pcif H[x]=\perp\\
    \pcind H[x]\leftarrow\mathcal{H}^{\mathsf{SQIsignHD}}(x)\\
    \pcreturn H[x]
    }
    \procedure[linenumbering]{$\mathcal{O}_{pS} (m, I_w)$}{%
    \text{Parse } I_w \text{ as } (E_w,w(\frak{B}),\pi_w)\\
    w := \mathcal{E}(E_w,w(\frak{B}), \pi_w , H)\\
    \pcif (w, I_w)\not\in \mathsf{R_{\mathfrak{A}}}\\
    \pcind \pcabort\\
    \mathsf{\Sigma}\leftarrow\mathsf{Sig}(\tau,m)\\
\text{Parse }\mathsf{\Sigma} \text{ as } (E_{1},\mathcal{R}_{\sigma})\\
\text{Extract} (E^{\ast}_{\psi},\mathcal{R}^{\ast}_{\tilde{\sigma}}) \text{ by } \mathcal{A}_{\mathsf{SIDH}} \text{ s.t. }\\
\pcind \sigma = \tilde{\sigma}\circ \hat{w}^{\ast} \text{ with } \deg(\hat{w}^{\ast})=C\\
\text{Extract }\alpha \text{ from }\\
\pcind \ker(w)=\langle P+[\alpha]Q\rangle\\
\text{Find } \langle P^{\ast},Q^{\ast}\rangle=E^{\ast}_{\psi}[C] \text{ for which}\\
\pcind \ker(w^{\ast})=\langle P^{\ast}+[\alpha]Q^{\ast}\rangle\\
%\pcind \,e_{C}(P,Q)=e_{C}(P^{\ast},Q^{\ast})^{B}\\
\text{Set } S^{\ast}:=(P^{\ast},Q^{\ast})\\
\pi^{\ast}_{\psi'}\leftarrow \mathcal{S}((E_w, E_1),1)\\
\mathcal{Q}:=\mathcal{Q}\cup\{m\}\\
\pcreturn \tilde{\mathsf{\Sigma}}:=(E_1, \pi^{\ast}_{\psi'}, E^{\ast}_{\psi},S^{\ast},\mathcal{R}^{\ast}_{\tilde{\sigma}})
    }
    \procedure[linenumbering]{$\mathcal{O}_S (m)$}{%
    \mathsf{\Sigma}\leftarrow\mathsf{Sig}(\tau,m)\\
    \mathcal{Q}:=\mathcal{Q}\cup\{m\} \\
    \pcreturn \mathsf{\Sigma}
    }
    \end{pcvstack}}
\end{pchstack}

\noindent $\mathsf{Game_4}$. In this game, upon receiving the challenge message $m^\ast$ from $\mathcal{A}$, the game itself generates a signature $\mathsf{\Sigma}$ by running the $\mathsf{Sig}$ algorithm and converting the resulting signature into a valid pre-signature, as in the previous game during the $\mathcal{O}_{pS}$ execution. Consequently, the same indistinguishability argument as in the previous game also holds. Therefore, it follows that 
$$Pr[\mathsf{Game_4} = 1] \leq Pr[\mathsf{Game_3} = 1] + \mathsf{negl}(\lambda).$$

\noindent After establishing that the transition from the original \textsf{aSigForge} game ($\mathsf{Game_0}$) to $\mathsf{Game_4}$ is indistinguishable, it remains to demonstrate the existence of a simulator that perfectly emulates $\mathsf{Game_4}$ and leverages $\mathcal{A}$ to succeed in the \textsf{StrongSigForge} game. Below, we provide a concise description of how the simulator responds to Oracle queries.\\

\noindent\textbf{Simulation of Oracle Queries:}\\

\noindent\textbf{Signing queries.} If the adversary $\mathcal{A}$ queries the signing oracle $\mathcal{O}_{S}$ on input $m$, $\mathcal{S}$ sends $m$ to its oracle $\mathsf{Sig}^{\mathsf{SQIsignHD}}$ and forwards its response to $\mathcal{A}$.

\noindent\textbf{Random Oracle queries.} Based on $\mathcal{A}$ querying the oracle $\mathcal{H}$ on input $x$, in case $H[x] =\perp$, then $\mathcal{S}$ queries $\mathcal{H}^\mathsf{SQIsignHD}(x)$, otherwise the simulator outputs $H[x]$.

\noindent\textbf{Pre-Signing queries.} If $\mathcal{A}$ queries the pre-signing oracle $\mathcal{O}_{pS}$ on input $(m, I_w)$:
\begin{enumerate}
    \item The simulator extracts the witness isogeny $w$ using the extractability property of $\mathsf{NIZK}$. It then forwards the message $m$ to the oracle $\mathsf{Sig}^{\mathsf{SQIsignHD}}$ and parses the generated signature $\mathsf{\Sigma}$ as $(E_1, \mathcal{R}_{\sigma})$.
    \item The simulator $\mathcal{S}$ constructs a pre-signature isogeny representation $\mathcal{R}_{\tilde{\sigma}}$, and torsion basis $S^{\ast}=(P^{\ast},Q^{\ast})$ by decomposing  $\sigma$ into $\tilde{\sigma}\circ\hat{w}^{\ast}$ using the algorithm $\mathcal{A}_{\mathsf{SIDH}}$, and the $\alpha$ which is obtained from the extracted witness $w:E_0 \to E_{0}/\langle P + [\alpha]Q\rangle$ via the online extractor property, respectively.
    \item Finally, $\mathcal{S}$ simulates a zero-knowledge proof $\pi^{\ast}_{\psi'}$, for the statement $E_1$. The simulator outputs $\tilde{\mathsf{\Sigma}} = (E_1, \pi^{\ast}_{\psi'}, E^{\ast}_{\psi}, S^{\ast}, \mathcal{R}^{\ast}_{\tilde{\sigma}})$.
\end{enumerate}

\noindent\textbf{Challenge phase:}
\begin{enumerate}
\item When $\mathcal{A}$ outputs the message $m^\ast$ as the challenge message, $\mathcal{S}$ generates $(w, I_w) \leftarrow \mathsf{GenR}(1^\lambda)$, forwards $m^\ast$ to the oracle $\mathsf{Sig}^{\mathsf{SQIsignHD}}$, and parses the generated signature $\mathsf{\Sigma}$ as $(E_1, \mathcal{R}_{\sigma})$.
\item $\mathcal{S}$ generates the required pre-signature $\tilde{\mathsf{\Sigma}}$ in the same manner as it does during $\mathcal{O}_{pS}$ queries.
\item When $\mathcal{A}$ outputs a forgery $\mathsf{\Sigma}^\ast$, the simulator outputs $(m^\ast, \mathsf{\Sigma}^\ast)$ as its own forgery.
\end{enumerate}

\noindent We highlight that the primary difference between the simulation and $\mathsf{Game_4}$ is syntactical. Specifically, rather than generating the secret/public keys and executing the algorithms $\mathsf{Sig}$ and $\mathcal{H}$, the simulator $\mathcal{S}$ utilizes its oracles $\mathsf{Sig}^{\mathsf{SQIsignHD}}$ and $\mathcal{H}^{\mathsf{SQIsignHD}}$. It remains to demonstrate that the forgery produced by $\mathcal{A}$ can be used by the simulator to win the $\mathsf{StrongSigForge}$ game.

\begin{claim}
    $(m^\ast,\mathsf{\Sigma}^\ast)$ constitutes a valid forgery in the $\mathsf{StrongSigForge}$ game.
    \begin{proof}
        To prove this claim, we must show that the pair $(m^\ast,\mathsf{\Sigma}^\ast)$ has not been previously output by the oracle $\mathsf{Sig^{SQIsignHD}}$. Note that the adversary $\mathcal{A}$ has not made a query on the challenge message $m^\ast$ to either $\mathcal{O}_{S}$ or $\mathcal{O}_{pS}$. Therefore, $\mathsf{Sig^{SQIsignHD}}$ is only queried on $m^\ast$ during the challenge phase. As demonstrated in the game $\mathsf{Game_1}$, the adversary produces a forgery $\mathsf{\Sigma}^\ast$, which matches the signature $\mathsf{\Sigma}$ output by $\mathsf{Sig^{SQIsignHD}}$ during the challenge phase with only negligible probability. Consequently, the oracle $\mathsf{Sig^{SQIsignHD}}$ has never output $\mathsf{\Sigma}^\ast$ on the query $m^\ast$ before, establishing that $(m^\ast, \mathsf{\Sigma}^\ast)$ is a valid forgery for the $\mathsf{StrongSigForge}$ game.
    \end{proof}
\end{claim}

    \begin{pchstack}[boxed, center, space=.3em]\label{fig:Simulator1}
    \begin{pcvstack}[space=.3em]
    \procedure[linenumbering]{$\mathcal{S}^{{\mathsf{Sig^{SQIsignHD}}},{\mathsf{\mathcal{H}^{SQIsignHD}}}}(E_\tau)$}{%
    \mathcal{Q}:=\emptyset \\
    H:=[\perp]\\
    (\tau,E_{\tau})\leftarrow \mathsf{KeyGen}(\secparam)\\
    m^\ast\leftarrow \adv^{\mathcal{H},\mathcal{O}_{S},\mathcal{O}_{pS}}(E_\tau)\\
    (w,I_w)\leftarrow \mathsf{GenR(1^{\lambda})}\\
\gamechange{$\mathsf{\Sigma}\leftarrow\mathsf{Sig^{SQIsignHD}}\big(m^\ast)$}\\
    \text{Parse }\mathsf{\Sigma} \text{ as } (E_{1},\mathcal{R}_{\sigma})\\
\text{Extract} (E^{\ast}_{\psi},\mathcal{R}^{\ast}_{\tilde{\sigma}}) \text{ by } \mathcal{A}_{\mathsf{SIDH}} \text{ s.t. }\\
\pcind \sigma = \tilde{\sigma}\circ \hat{w}^{\ast} \text{ with } \deg(\hat{w}^{\ast})=C\\
\text{Extract }\alpha \text{ from }\\
\pcind \ker(w)=\langle P+[\alpha]Q\rangle\\
\text{Find } \langle P^{\ast},Q^{\ast}\rangle=E^{\ast}_{\psi}[C] \text{ for which}\\
\pcind \ker(w^{\ast})=\langle P^{\ast}+[\alpha]Q^{\ast}\rangle\\
%\pcind \,e_{C}(P,Q)=e_{C}(P^{\ast},Q^{\ast})^{B}\\
\text{Set } S^{\ast}:=(P^{\ast},Q^{\ast})\\
\pi^{\ast}_{\psi'}\leftarrow \mathcal{S}((E_w, E_1),1)\\
\tilde{\mathsf{\Sigma}}:=(E_1, \pi^{\ast}_{\psi'}, E^{\ast}_{\psi},S^{\ast},\mathcal{R}^{\ast}_{\tilde{\sigma}})\\
    \mathsf{\Sigma}^\ast\leftarrow \adv^{\mathcal{H},\mathcal{O}_{S},\mathcal{O}_{pS}}(\mathsf{\tilde{\Sigma}}, I_w)\\
    \pcreturn (m^\ast,\mathsf{\Sigma}^\ast)
    }
    \procedure[linenumbering]{$\mathcal{O}_S (m)$}{%
    \gamechange{$\mathsf{\Sigma}\leftarrow\mathsf{Sig}^{\mathsf{SQIsignHD}}(m)$}\\
    \mathcal{Q}:=\mathcal{Q}\cup\{m\} \\
    \pcreturn \mathsf{\Sigma}
    }
    \end{pcvstack}

    \begin{pcvstack}[space=.3em]
    
    \procedure[linenumbering]{$\mathcal{H} (x)$}{%
    \pcif H[x]=\perp\\
    \gamechange{$\pcind H[x]\leftarrow\mathcal{H}^{\mathsf{SQIsignHD}}(x)$}\\
    \pcreturn H[x]
    }
    \procedure[linenumbering]{$\mathcal{O}_{pS} (m,I_w)$}{%
    \text{Parse } I_w \text{ as } (E_w,w(\frak{B}),\pi_w)\\
    w := \mathcal{E}(E_w,w(\frak{B}), \pi_w , H)\\
    \pcif (w, I_w)\not\in \mathsf{R_{\mathfrak{A}}}\\
    \pcind \pcabort\\
\mathsf{\Sigma}\leftarrow\mathsf{Sig}(\tau,m)\\
\text{Parse }\mathsf{\Sigma} \text{ as } (E_{1},\mathcal{R}_{\sigma})\\
\text{Extract} (E^{\ast}_{\psi},\mathcal{R}^{\ast}_{\tilde{\sigma}}) \text{ by } \mathcal{A}_{\mathsf{SIDH}} \text{ s.t. }\\
\pcind \sigma = \tilde{\sigma}\circ \hat{w}^{\ast} \text{ with } \deg(\hat{w}^{\ast})=C\\
\text{Extract }\alpha \text{ from }\\
\pcind \ker(w)=\langle P+[\alpha]Q\rangle\\
\text{Find } \langle P^{\ast},Q^{\ast}\rangle=E^{\ast}_{\psi}[C] \text{ for which}\\
\pcind \ker(w^{\ast})=\langle P^{\ast}+[\alpha]Q^{\ast}\rangle\\
%\pcind \,e_{C}(P,Q)=e_{C}(P^{\ast},Q^{\ast})^{B}\\
\text{Set } S^{\ast}:=(P^{\ast},Q^{\ast})\\
\pi^{\ast}_{\psi'}\leftarrow \mathcal{S}((E_w, E_1),1)\\
\mathcal{Q}:=\mathcal{Q}\cup\{m\}\\
\pcreturn \tilde{\mathsf{\Sigma}}:=(E_1, \pi^{\ast}_{\psi'}, E^{\ast}_{\psi},S^{\ast},\mathcal{R}^{\ast}_{\tilde{\sigma}})
    }
    \end{pcvstack}
\end{pchstack}

\noindent From the game $\mathsf{Game_0}$ to the game $\mathsf{Game_4}$, we have that 
$$\text{Pr}[\mathsf{Game_0} = 1] \leq \text{Pr}[\mathsf{Game_4} = 1] + \mathsf{negl}(\lambda).$$ 
Due to a perfect simulation of $\mathsf{Game_4}$, provided by the simulator $\mathcal{S}$, it follows that
\begin{align*}
    \mathsf{Adv}_{\mathcal{A}}^{\mathsf{aSigForge}} &= \text{Pr}[\mathsf{Game_0} = 1]\\
    &\leq \text{Pr}[\mathsf{Game_4} = 1] + \mathsf{negl}(\lambda)\\
    &\leq \mathsf{Adv}_{\mathcal{S}}^{\mathsf{StrongSigForge}} + \mathsf{negl}(\lambda).
\end{align*}

\noindent By assumption, as SQIsignHD is secure in ROM with $\mathcal{H}^{\mathsf{SQIsignHD}}$ programmed as a random oracle, it implies that our adaptor signature, $\Xi_\mathsf{R_{\mathfrak{A}},\Sigma_{SQIsignHD}}$, is $\mathsf{aEUF\text{-}CMA}$ secure in ROM. This completes the proof of Lemma \ref{lemma:securityLemma3}.
\end{proof}
\end{lemma}

\begin{lemma}\label{lemma:securityLemma4}
    Assuming that the SQIsignHD signature scheme $\mathsf{\Sigma_{SQIsignHD}}$ is $\mathsf{SUF}$-$\mathsf{CMA}$-secure, that $\mathsf{R_{\mathfrak{A}}}$ is a hard relation, and that Problem \ref{problem:SSIP-A} and Problem \ref{problem:SSIP-B} are computationally hard, then the $\mathsf{SQIAsignHD}$ adaptor signature scheme $\Xi_\mathsf{R_{\mathfrak{A}},\Sigma_{SQIsignHD}}$, as given in Algorithm \ref{alg:SQIAsignHD}, is witness extractable.
    \begin{proof}
    We begin by outlining the primary intuition behind the proof of witness extractability. The proof of this lemma closely follows the proof of Lemma  \ref{lemma:securityLemma3}. Specifically, we prove this lemma by reducing the witness extractability of $\Xi_\mathsf{R_{\mathfrak{A}},\Sigma_{SQIsignHD}}$ to the strong unforgeability of the SQIsignHD signature scheme, $\mathsf{\Sigma_{SQIsignHD}}$. To do so, let $\mathcal{A}$ be a $\mathsf{PPT}$ adversary that wins the $\mathsf{aWitExt}$ game. We then construct another $\mathsf{PPT}$ adversary, $\mathcal{S}$, so that it wins the $\mathsf{StrongSigForge}$ game.

    \noindent Analogous to the proof of  Lemma \ref{lemma:securityLemma3}, the primary challenge lies in simulating the pre-signing queries. Consequently, the simulation process is carried out exactly as in the proof of Lemma \ref{lemma:securityLemma3}. 
    
    The key distinction in this case, however, is that in the $\mathsf{aWitExt}$ game, the adversary $\mathcal{A}$ outputs the statement $I_w$ for the relation $\mathsf{R_{\mathfrak{A}}}$ along with the challenge message $m^\ast$. This implies that the pair $(w, I_w)$ is not predetermined by the game. As a result, $\mathcal{S}$ cannot convert a valid signature into a pre-signature since it does not have access to the witness $w$. However, $w$ can be extracted from the zero-knowledge proof embedded in $I_w$. Once $w$ is extracted, then $S$ can simulate the pre-signing queries as in Lemma \ref{lemma:securityLemma3}. We, now, begin with designing a series of games required for the proof.\\

    \noindent $\mathsf{Game_0}$. This game corresponds to the original $\mathsf{aWitExt}$ game, as per  Definition \ref{def:aWitExtGame}, where the adversary $\mathcal{A}$ must produce a valid signature $\mathsf{\Sigma}$ for a message $m$ of its choice, given a pre-signature $\tilde{\mathsf{\Sigma}}$ and a witness/statement pair $(w,I_w)$, while having access to the oracles $\mathcal{H}$, $\mathcal{O}_{pS}$ and $\mathcal{O}_{S}$. The adversary $\mathcal{A}$ succeeds if $(\mathsf{Ext}(\tilde{\mathsf{\Sigma}}, \mathsf{\Sigma}, I_w),I_w)\not\in\mathsf{R_{\mathfrak{A}}}$. Since we are in the random oracle model, we explicitly write the random oracle code $\mathcal{H}$. It then trivially follows that:
    $$\text{Pr}[\mathsf{Game_0}=1]=\text{Pr}[\mathsf{aWitExt}_{{\mathcal{A},\Xi_{\mathsf{R_{\frak{A}},\mathsf{\Sigma_{SQIsignHD}}}}}}(\lambda)=1].$$
    
    \begin{pchstack}[boxed, center, space=0.3em] \label{fig:Witgame0}
    \begin{pcvstack}[space=.3em]
    \procedure[linenumbering]{\pcgame [0]}{%
    \mathcal{Q}:=\emptyset \\
    H:=[\perp]\\
    (\tau,E_\tau)\leftarrow \mathsf{KeyGen}(\secparam)\\
    (m^{\ast},I_w)\leftarrow \adv^{\mathcal{H},\mathcal{O}_{S},\mathcal{O}_{pS}}(E_\tau)\\
    \tilde{\mathsf{\Sigma}}\leftarrow\mathsf{PreSig}(\tau,m^{\ast},I_w)\\
    \mathsf{\Sigma}^{\ast}\leftarrow\adv^{\mathcal{H},\mathcal{O}_{S},\mathcal{O}_{pS}}(\tilde{\mathsf{\Sigma}})\\
    w^\ast :=\mathsf{Ext}(\tilde{\mathsf{\Sigma}}, \mathsf{\Sigma}^{\ast},I_w)\\
    b_1 :=\mathsf{Ver}(E_\tau,m^{\ast},\mathsf{\Sigma}^{\ast})\\
    b_2 := m^{\ast}\not\in \mathcal{Q}\\
    b_3 :=(w^\ast, I_w)\not\in \mathsf{R_{\mathfrak{A}}} \\
    \pcreturn b_1\land b_2\land b_3
    }
    \end{pcvstack}

    \begin{pcvstack}[space=.3em]
    
    \procedure[linenumbering]{$\mathcal{H} (x)$}{%
    \pcif H[x]=\perp\\
    \pcind H[x]\leftarrow\mathcal{H}^{\mathsf{SQIsignHD}}(x)\\
    \pcreturn H[x]
    }
    \procedure[linenumbering]{$\mathcal{O}_{pS}(m,I_w)$}{%
    \tilde{\mathsf{\Sigma}}\leftarrow\mathsf{PreSig}(\tau,m,I_w)\\
    \mathcal{Q}:=\mathcal{Q}\cup\{m\}\\
    \pcreturn \tilde{\mathsf{\Sigma}}
    }
    \procedure[linenumbering]{$\mathcal{O}_S \big(m)$}{%
    \mathsf{\Sigma}\leftarrow\mathsf{Sig}(\tau, m) \\
    \mathcal{Q}:=\mathcal{Q}\cup\{m\} \\
    \pcreturn \mathsf{\Sigma}
    }
    \end{pcvstack}
\end{pchstack}

\noindent $\mathsf{Game_1}$. This game is the same as $\mathsf{Game_0}$, except that some changes are applied to the pre-signing oracle $\mathcal{O}_{pS}$. More specifically, during the $\mathcal{O}_{pS}$ queries, this game extracts a witness $w$ by executing the online extractor algorithm $\mathcal{E}$ on the inputs: the statement $(E_w,w(\frak{B}))$, the proof $\pi_w$, and the list of random oracle queries $H$. The game aborts if the extracted witness $w$ does not satisfy $(w, I_w)\in \mathsf{R_{\mathfrak{A}}}$.

    \begin{pchstack}[boxed, center, space=0.3em] \label{fig:Witgame1}
    \begin{pcvstack}[space=.3em]
    \procedure[linenumbering]{\pcgame [1]}{%
    \mathcal{Q}:=\emptyset \\
    H:=[\perp]\\
    (\tau,E_\tau)\leftarrow \mathsf{KeyGen}(\secparam)\\
    (m^{\ast},I_w)\leftarrow \adv^{\mathcal{H},\mathcal{O}_{S},\mathcal{O}_{pS}}(E_\tau)\\
    \tilde{\mathsf{\Sigma}}\leftarrow\mathsf{PreSig}(\tau,m^{\ast},I_w)\\
    \mathsf{\Sigma}^{\ast}\leftarrow\adv^{\mathcal{H},\mathcal{O}_{S},\mathcal{O}_{pS}}(\tilde{\mathsf{\Sigma}})\\
    w^\ast :=\mathsf{Ext}(\tilde{\mathsf{\Sigma}}, \mathsf{\Sigma}^{\ast},I_w)\\
    b_1 :=\mathsf{Ver}(E_\tau,m^{\ast},\mathsf{\Sigma}^{\ast})\\
    b_2 := m^{\ast}\not\in \mathcal{Q}\\
    b_3 :=(w^\ast, I_w)\not\in \mathsf{R_{\mathfrak{A}}} \\
    \pcreturn b_1\land b_2\land b_3
    }
     \procedure[linenumbering]{$\mathcal{O}_S \big(m)$}{%
    \mathsf{\Sigma}\leftarrow\mathsf{Sig}(\tau, m) \\
    \mathcal{Q}:=\mathcal{Q}\cup\{m\} \\
    \pcreturn \mathsf{\Sigma}
    }
    \end{pcvstack}

    \begin{pcvstack}[space=.3em]
    
    \procedure[linenumbering]{$\mathcal{H} (x)$}{%
    \pcif H[x]=\perp\\
    \pcind H[x]\leftarrow\mathcal{H}^{\mathsf{SQIsignHD}}(x)\\
    \pcreturn H[x]
    }
    \procedure[linenumbering]{$\mathcal{O}_{pS}(m,I_w)$}{%
    \gamechange{Parse $I_w$ as $(E_w,w(\frak{B}),\pi_w)$}\\
    \gamechange{$w := \mathcal{E}(E_w,w(\frak{B}),\pi_w, H)$}\\
    \gamechange{$\pcif (w,I_w)\not\in \mathsf{R_{\mathfrak{A}}}$}\\
    \pcind\gamechange{$\pcabort$}\\
    \tilde{\mathsf{\Sigma}}\leftarrow\mathsf{PreSig}(\tau,m,I_w)\\
    \mathcal{Q}:=\mathcal{Q}\cup\{m\}\\
    \pcreturn \tilde{\mathsf{\Sigma}}
    }
    \end{pcvstack}
\end{pchstack}

\begin{claim}\label{claim:witness_bad1}
    If $\mathsf{Bad_1}$ is the event that $\mathsf{Game_1}$ aborts while the execution of $\mathcal{O}_{pS}$, then it holds that Pr$[\mathsf{Bad_1}] \leq \negl$.
    \begin{proof}
        By the online extractor property of $\mathsf{NIZK}$, if a witness $w$ is extracted from a proof $\pi_{w}$ for the statement $(E_w, w(\mathfrak{B}))$ such that $\mathsf{NIZK.V}((E_w, w(\mathfrak{B})), \pi_w) = 1$, it follows that $(w, I_w) \in \mathsf{R_{\mathfrak{A}}}$, except with negligible probability.
    \end{proof}
\end{claim}
It follows that $\mathsf{Game_1}$ and $\mathsf{Game_0}$ are equivalent, except when the event $\mathsf{Bad_1}$ occurs. Thus, we have:
$$\text{Pr}[\mathsf{Game_0} = 1] \leq \text{Pr}[\mathsf{Game_1} = 1] + \mathsf{negl}(\lambda).$$

\begin{pchstack}[boxed, center, space=0.3em] \label{fig:Witgame2}
    \begin{pcvstack}[space=.3em]
    \procedure[linenumbering]{\pcgame [2]}{%
    \mathcal{Q}:=\emptyset \\
    H:=[\perp]\\
    (\tau,E_\tau)\leftarrow \mathsf{KeyGen}(\secparam)\\
    (m^{\ast},I_w)\leftarrow \adv^{\mathcal{H},\mathcal{O}_{S},\mathcal{O}_{pS}}(E_\tau)\\
    \tilde{\mathsf{\Sigma}}\leftarrow\mathsf{PreSig}(\tau,m^{\ast},I_w)\\
    \mathsf{\Sigma}^{\ast}\leftarrow\adv^{\mathcal{H},\mathcal{O}_{S},\mathcal{O}_{pS}}(\tilde{\mathsf{\Sigma}})\\
    w^\ast :=\mathsf{Ext}(\tilde{\mathsf{\Sigma}}, \mathsf{\Sigma}^{\ast},I_w)\\
    b_1 :=\mathsf{Ver}(E_\tau,m^{\ast},\mathsf{\Sigma}^{\ast})\\
    b_2 := m^{\ast}\not\in \mathcal{Q}\\
    b_3 :=(w^\ast, I_w)\not\in \mathsf{R_{\mathfrak{A}}} \\
    \pcreturn b_1\land b_2\land b_3
    }
     \procedure[linenumbering]{$\mathcal{O}_S \big(m)$}{%
    \mathsf{\Sigma}\leftarrow\mathsf{Sig}(\tau, m) \\
    \mathcal{Q}:=\mathcal{Q}\cup\{m\} \\
    \pcreturn \mathsf{\Sigma}
    }
    \end{pcvstack}

    \begin{pcvstack}[space=.3em]
    
    \procedure[linenumbering]{$\mathcal{H} (x)$}{%
    \pcif H[x]=\perp\\
    \pcind H[x]\leftarrow\mathcal{H}^{\mathsf{SQIsignHD}}(x)\\
    \pcreturn H[x]
    }
    \procedure[linenumbering]{$\mathcal{O}_{pS} (m,I_w)$}{%
    \text{Parse } I_w \text{ as } (E_w,w(\frak{B}),\pi_w)\\
    w := \mathcal{E}(E_w,w(\frak{B}), \pi_w , H)\\
    \pcif (w, I_w)\not\in \mathsf{R_{\mathfrak{A}}}\\
    \pcind \pcabort\\
\gamechange{$\mathsf{\Sigma}\leftarrow\mathsf{Sig}(\tau,m)$}\\
\gamechange{$\text{Parse }\mathsf{\Sigma} \text{ as } (E_{1},\mathcal{R}_{\sigma})$}\\
\gamechange{\text{Extract} $(E^{\ast}_{\psi},\mathcal{R}^{\ast}_{\tilde{\sigma}}) \text{ by } \mathcal{A}_{\mathsf{SIDH}} \text{ s.t. }$}\\
\pcind \gamechange{$\sigma = \tilde{\sigma}\circ \hat{w}^{\ast}$ with $\deg(\hat{w}^{\ast})=C$}\\
\gamechange{$\text{Extract }\alpha \text{ from }$}\\
\pcind \gamechange{$\ker(w)=\langle P+[\alpha]Q\rangle$}\\
\gamechange{Find $\langle P^{\ast},Q^{\ast}\rangle=E^{\ast}_{\psi}[C]$ for which}\\
\pcind\gamechange{$\ker(w^{\ast})=\langle P^{\ast}+[\alpha]Q^{\ast}\rangle$}\\
%\pcind\gamechange{$e_{C}(P,Q)=e_{C}(P^{\ast},Q^{\ast})^{B}$}\\
\gamechange{Set $S^{\ast}:=(P^{\ast},Q^{\ast})$}\\
\gamechange{$\pi^{\ast}_{\psi'}\leftarrow \mathcal{S}((E_w, E_1),1)$}\\
\mathcal{Q}:=\mathcal{Q}\cup\{m\}\\
\gamechange{$\pcreturn \tilde{\mathsf{\Sigma}}:=(E_1, \pi^{\ast}_{\psi'}, E^{\ast}_{\psi},S^{\ast},\mathcal{R}^{\ast}_{\tilde{\sigma}})$}
    }
    \end{pcvstack}
\end{pchstack}

\noindent $\mathsf{Game_2}$. This game extends the modifications to the pre-signing oracle $\mathcal{O}_{pS}$ from the previous game. It generates a valid signature $\mathsf{\Sigma} = (E_1, \mathcal{R}_{\sigma})$ using the $\mathsf{Sig}$ algorithm and decomposes the isogeny $\sigma$ into $\sigma = \tilde{\sigma} \circ \hat{w}^{\ast}$ using $\mathcal{A}_{\mathsf{SIDH}}$. Here, $\hat{w}^{\ast}$ is a $C$-isogeny from $E_1$ to a curve $E^{\ast}_{\psi}$, and $\tilde{\sigma}$ is an isogeny from $E^{\ast}_{\psi}$ to $E_2$. Thereby, the efficient representation $\mathcal{R}_{\tilde{\sigma}}$ for $\tilde{\sigma}$ is computed.

To construct a $C$-torsion basis $\langle P^{\ast}, Q^{\ast} \rangle = E_{\psi}^{\ast}[C]$ for which $\ker(w^{\ast}) = \langle P^{\ast} + [\alpha]Q^{\ast} \rangle$ with $\alpha$ derived from the witness $w$, a basis $\langle R, R' \rangle = E_{\psi}^{\ast}[C]$ is formed by selecting a point $R'$ linearly independent of $R = \ker(w^{\ast})$. The coefficients $x_1, y_1, x_2, y_2$ are determined, where $(x_1, x_2), (y_1, y_2) \in (\mathbb{Z}/C\mathbb{Z}) \times (\mathbb{Z}/C\mathbb{Z})$ and $(x_i, y_i) \neq (0, 0)$ for $i = 1, 2$,  to satisfy:

$$(x_1R + y_1R’) + \alpha(x_2R + y_2R’) = R.$$
These coefficients define the $C$-torsion basis for $E_{\psi}^{\ast}$ by setting $P^{\ast} = x_1R + y_1R'$ and $Q^{\ast} = x_2R + y_2R'$.

Finally, the simulator $\mathcal{S}$ generates a proof $\pi^{\ast}_{\psi'}$ for $E_{1}$ without knowledge of the isogeny $\psi'$, that is computationally indistinguishable from an honest proof, and the pre-signature is constructed as $\tilde{\mathsf{\Sigma}} = (E_1, \pi^{\ast}_{\psi'}, E^{\ast}_{\psi}, S^{\ast}, \mathcal{R}^{\ast}_{\tilde{\sigma}})$. The game remains indistinguishable from the previous one, ensuring:
\[
\text{Pr}[\mathsf{Game_1} = 1] \leq \text{Pr}[\mathsf{Game_2} = 1] + \mathsf{negl}(\lambda).
\]

\begin{pchstack}[boxed, center, space=0.3em] \label{fig:Witgame3}
    \begin{pcvstack}[space=.3em]
    \procedure[linenumbering]{\pcgame [3]}{%
    \mathcal{Q}:=\emptyset \\
    H:=[\perp]\\
    (\tau,E_\tau)\leftarrow \mathsf{KeyGen}(\secparam)\\
    (m^{\ast},I_w)\leftarrow \adv^{\mathcal{H},\mathcal{O}_{S},\mathcal{O}_{pS}}(E_\tau)\\
    \gamechange{Parse $I_w$ as $(E_w,w(\frak{B}),\pi_w)$}\\
    \gamechange{$w := \mathcal{E}(E_w,w(\frak{B}),\pi_w, H)$}\\
    \gamechange{$\pcif (w,I_w)\not\in \mathsf{R_{\mathfrak{A}}}$}\\
    \pcind\gamechange{$\pcabort$}\\
    \tilde{\mathsf{\Sigma}}\leftarrow\mathsf{PreSig}(\tau,m^{\ast},I_w)\\
    \mathsf{\Sigma}^{\ast}\leftarrow\adv^{\mathcal{H},\mathcal{O}_{S},\mathcal{O}_{pS}}(\tilde{\mathsf{\Sigma}})\\
    w^\ast :=\mathsf{Ext}(\tilde{\mathsf{\Sigma}}, \mathsf{\Sigma}^{\ast},I_w)\\
    b_1 :=\mathsf{Ver}(E_\tau,m^{\ast},\mathsf{\Sigma}^{\ast})\\
    b_2 := m^{\ast}\not\in \mathcal{Q}\\
    b_3 :=(w^\ast, I_w)\not\in \mathsf{R_{\mathfrak{A}}} \\
    \pcreturn b_1\land b_2\land b_3
    }
     \procedure[linenumbering]{$\mathcal{O}_S \big(m)$}{%
    \mathsf{\Sigma}\leftarrow\mathsf{Sig}(\tau, m) \\
    \mathcal{Q}:=\mathcal{Q}\cup\{m\} \\
    \pcreturn \mathsf{\Sigma}
    }
    \end{pcvstack}

    \begin{pcvstack}[space=.3em]
    
    \procedure[linenumbering]{$\mathcal{H} (x)$}{%
    \pcif H[x]=\perp\\
    \pcind H[x]\leftarrow\mathcal{H}^{\mathsf{SQIsignHD}}(x)\\
    \pcreturn H[x]
    }
    \procedure[linenumbering]{$\mathcal{O}_{pS} (m,I_w)$}{%
    \text{Parse } I_w \text{ as } (E_w,w(\frak{B}),\pi_w)\\
    w := \mathcal{E}(E_w,w(\frak{B}), \pi_w , H)\\
    \pcif (w, I_w)\not\in \mathsf{R_{\mathfrak{A}}}\\
    \pcind \pcabort\\
\mathsf{\Sigma}\leftarrow\mathsf{Sig}(\tau,m)\\
\text{Parse }\mathsf{\Sigma} \text{ as } (E_{1},\mathcal{R}_{\sigma})\\
\text{Extract } (E^{\ast}_{\psi},\mathcal{R}^{\ast}_{\tilde{\sigma}}) \text{ by } \mathcal{A}_{\mathsf{SIDH}} \text{ s.t. }\\
\pcind \sigma = \tilde{\sigma}\circ \hat{w}^{\ast} \text{ with } \deg(\hat{w}
^{\ast})=C\\
\text{Extract }\alpha \text{ from }\\
\pcind \ker(w)=\langle P+[\alpha]Q\rangle\\
\text{Find } \langle P^{\ast},Q^{\ast}\rangle=E^{\ast}_{\psi}[C] \text{ for which}\\
\pcind \ker(w^{\ast})=\langle P^{\ast}+[\alpha]Q^{\ast}\rangle\\
%\pcind\gamechange{$e_{C}(P,Q)=e_{C}(P^{\ast},Q^{\ast})^{B}$}\\
\text{Set } S^{\ast}:=(P^{\ast},Q^{\ast})\\
\pi^{\ast}_{\psi'}\leftarrow \mathcal{S}((E_w, E_1),1)\\
\mathcal{Q}:=\mathcal{Q}\cup\{m\}\\
\pcreturn \tilde{\mathsf{\Sigma}}:=(E_1, \pi^{\ast}_{\psi'}, E^{\ast}_{\psi},S^{\ast},\mathcal{R}^{\ast}_{\tilde{\sigma}})
    }
    \end{pcvstack}
\end{pchstack}

\noindent $\mathsf{Game_3}$. In this game, for the challenge phase, we apply the identical modifications implemented in $\mathsf{Game_1}$'s $\mathcal{O}_{pS}$ oracle. In the challenge phase, a witness $w$ is extracted by the online extractor algorithm $\mathcal{E}$ taking the statement $(E_w,w(\frak{B}))$, the proof $\pi_w$, and the list of random oracle queries $H$ as inputs. In case for the extracted witness $w$, the relation $\big(w,I_w)\in \mathsf{R_{\mathfrak{A}}}$ is not satisfied, then the game aborts.

\begin{claim}
    If $\mathsf{Bad_2}$ is the event that $\mathsf{Game_3}$ aborts during the challenge phase, then it follows Pr$[\mathsf{Bad_2}] \leq \negl$.
    \begin{proof}
The arguments presented in the proof of Claim \ref{claim:witness_bad1} apply similarly to prove this claim.
    \end{proof}
\end{claim}
Hence, $\mathsf{Game_3}$ and $\mathsf{Game_2}$ are equivalent except for the case that the event $\mathsf{Bad_2}$ happens. Thus, we have
$$\text{Pr}[\mathsf{Game_2} = 1] \leq \text{Pr}[\mathsf{Game_3} = 1] + \mathsf{negl}(\lambda).$$

\begin{pchstack}[boxed, center, space=0.3em] \label{fig:Witgame4}
    \begin{pcvstack}[space=.3em]
    \procedure[linenumbering]{\pcgame [4]}{%
    \mathcal{Q}:=\emptyset \\
    H:=[\perp]\\
    (\tau,E_\tau)\leftarrow \mathsf{KeyGen}(\secparam)\\
    (m^{\ast},I_w)\leftarrow \adv^{\mathcal{H},\mathcal{O}_{S},\mathcal{O}_{pS}}(E_\tau)\\
    \text{Parse } I_w \text{ as } (E_w,w(\frak{B}),\pi_w)\\
    w := \mathcal{E}(E_w,w(\frak{B}), \pi_w , H)\\
    \pcif (w, I_w)\not\in \mathsf{R_{\mathfrak{A}}}\\
    \pcind \pcabort\\
    \gamechange{$\mathsf{\Sigma}\leftarrow\mathsf{Sig}(\tau,m)$}\\
\gamechange{$\text{Parse }\mathsf{\Sigma} \text{ as } (E_{1},\mathcal{R}_{\sigma})$}\\
\gamechange{\text{Extract} $(E^{\ast}_{\psi},\mathcal{R}^{\ast}_{\tilde{\sigma}}) \text{ by } \mathcal{A}_{\mathsf{SIDH}} \text{ s.t. }$}\\
\pcind \gamechange{$\sigma = \tilde{\sigma}\circ \hat{w}^{\ast}$ with $\deg(\hat{w}^{\ast})=C$}\\
\gamechange{$\text{Extract }\alpha \text{ from }$}\\
\pcind \gamechange{$\ker(w)=\langle P+[\alpha]Q\rangle$}\\
\gamechange{Find $\langle P^{\ast},Q^{\ast}\rangle=E^{\ast}_{\psi}[C]$ for which}\\
\pcind\gamechange{$\ker(w^{\ast})=\langle P^{\ast}+[\alpha]Q^{\ast}\rangle$}\\
%\pcind\gamechange{$e_{C}(P,Q)=e_{C}(P^{\ast},Q^{\ast})^{B}$}\\
\gamechange{Set $S^{\ast}:=(P^{\ast},Q^{\ast})$}\\
\gamechange{$\pi^{\ast}_{\psi'}\leftarrow \mathcal{S}((E_w, E_1),1)$}\\
\gamechange{$\tilde{\mathsf{\Sigma}}:=(E_1, \pi^{\ast}_{\psi'}, E^{\ast}_{\psi},S^{\ast},\mathcal{R}^{\ast}_{\tilde{\sigma}})$}\\
\mathsf{\Sigma}^{\ast}\leftarrow\adv^{\mathcal{H},\mathcal{O}_{S},\mathcal{O}_{pS}}(\tilde{\mathsf{\Sigma}})\\
    w^\ast :=\mathsf{Ext}(\tilde{\mathsf{\Sigma}}, \mathsf{\Sigma}^{\ast},I_w)\\
    b_1 :=\mathsf{Ver}(E_\tau,m^{\ast},\mathsf{\Sigma}^{\ast})\\
    b_2 := m^{\ast}\not\in \mathcal{Q}\\
    b_3 :=(w^\ast, I_w)\not\in \mathsf{R_{\mathfrak{A}}} \\
    \pcreturn b_1\land b_2\land b_3
    }
    \end{pcvstack}

    \begin{pcvstack}[space=.3em]
    
    \procedure[linenumbering]{$\mathcal{H} (x)$}{%
    \pcif H[x]=\perp\\
    \pcind H[x]\leftarrow\mathcal{H}^{\mathsf{SQIsignHD}}(x)\\
    \pcreturn H[x]
    }
    \procedure[linenumbering]{$\mathcal{O}_{pS} (m,I_w)$}{%
    \text{Parse } I_w \text{ as } (E_w,w(\frak{B}),\pi_w)\\
    w := \mathcal{E}(E_w,w(\frak{B}), \pi_w , H)\\
    \pcif (w, I_w)\not\in \mathsf{R_{\mathfrak{A}}}\\
    \pcind \pcabort\\
\mathsf{\Sigma}\leftarrow\mathsf{Sig}(\tau,m)\\
\text{Parse }\mathsf{\Sigma} \text{ as } (E_{1},\mathcal{R}_{\sigma})\\
\text{Extract } (E^{\ast}_{\psi},\mathcal{R}^{\ast}_{\tilde{\sigma}}) \text{ by } \mathcal{A}_{\mathsf{SIDH}} \text{ s.t. }\\
\pcind \sigma = \tilde{\sigma}\circ \hat{w}^{\ast} \text{ with } \deg(\hat{w}^{\ast})=C\\
\text{Extract }\alpha \text{ from }\\
\pcind \ker(w)=\langle P+[\alpha]Q\rangle\\
\text{Find } \langle P^{\ast},Q^{\ast}\rangle=E^{\ast}_{\psi}[C] \text{ for which}\\
\pcind \ker(w^{\ast})=\langle P^{\ast}+[\alpha]Q^{\ast}\rangle\\
%\pcind\gamechange{$e_{C}(P,Q)=e_{C}(P^{\ast},Q^{\ast})^{B}$}\\
\text{Set } S^{\ast}:=(P^{\ast},Q^{\ast})\\
\pi^{\ast}_{\psi'}\leftarrow \mathcal{S}((E_w, E_1),1)\\
\mathcal{Q}:=\mathcal{Q}\cup\{m\}\\
\pcreturn \tilde{\mathsf{\Sigma}}:=(E_1, \pi^{\ast}_{\psi'}, E^{\ast}_{\psi},S^{\ast},\mathcal{R}^{\ast}_{\tilde{\sigma}})
    }
    \procedure[linenumbering]{$\mathcal{O}_S \big(m)$}{%
    \mathsf{\Sigma}\leftarrow\mathsf{Sig}(\tau, m) \\
    \mathcal{Q}:=\mathcal{Q}\cup\{m\} \\
    \pcreturn \mathsf{\Sigma}
    }
    \end{pcvstack}
\end{pchstack}

\noindent $\mathsf{Game_4}$. In this game, the challenge phase employs similar modifications implemented in $\mathsf{Game_2}$ for the $\mathcal{O}_{pS}$ oracle. Specifically, the game begins by generating a valid full signature $\mathsf{\Sigma}$ using the $\mathsf{Sig}$ algorithm and subsequently converts $\mathsf{\Sigma}$ into a pre-signature with the help of the extracted witness $w$ and $\mathcal{A}_{\mathsf{SIDH}}$. Additionally, the game computes the zero-knowledge proof in the same manner as described in $\mathsf{Game_2}$. Consequently, the same arguments apply here as well. Thus, this game is indistinguishable from the previous one, and it follows that
$$\text{Pr}[\mathsf{Game_3} = 1] \leq \text{Pr}[\mathsf{Game_4} = 1] + \mathsf{negl}(\lambda).$$

\noindent Having established that the transformation of the original $\mathsf{aWitExt}$ game into $\mathsf{Game_4}$ is indistinguishable, it remains to demonstrate the existence of a simulator that perfectly simulates $\mathsf{Game_4}$ while leveraging the adversary $\mathcal{A}$ to win the $\mathsf{StrongSigForge}$ game. Below, we provide a concise description of the simulator’s implementation.

    \begin{pchstack}[boxed, center, space=0.3em] \label{fig:Simulator2}
    \begin{pcvstack}[space=.3em]
    \procedure[linenumbering]{$\mathcal{S}^{{\mathsf{Sig^{SQIsignHD}}},{\mathsf{\mathcal{H}^{SQIsignHD}}}}(E_\tau)$}{%
    \mathcal{Q}:=\emptyset \\
    H:=[\perp]\\
    (\tau,E_\tau)\leftarrow \mathsf{KeyGen}(\secparam)\\
    (m^{\ast},I_w)\leftarrow \adv^{\mathcal{H},\mathcal{O}_{S},\mathcal{O}_{pS}}(E_\tau)\\
    \text{Parse } I_w \text{ as } (E_w,w(\frak{B}),\pi_w)\\
    w := \mathcal{E}(E_w,w(\frak{B}), \pi_w , H)\\
    \pcif (w, I_w)\not\in \mathsf{R_{\mathfrak{A}}}\\
    \pcind \pcabort\\
    \gamechange{$\mathsf{\Sigma}\leftarrow\mathsf{Sig^{SQIsignHD}}(m)$}\\
\text{Parse }\mathsf{\Sigma} \text{ as } (E_{1},\mathcal{R}_{\sigma})\\
\text{Extract } (E^{\ast}_{\psi},\mathcal{R}^{\ast}_{\tilde{\sigma}}) \text{ by } \mathcal{A}_{\mathsf{SIDH}} \text{ s.t. }\\
\pcind \sigma = \tilde{\sigma}\circ \hat{w}^{\ast} \text{ with } \deg(\hat{w}^{\ast})=C\\
\text{Extract }\alpha \text{ from }\\
\pcind \ker(w)=\langle P+[\alpha]Q\rangle\\
\text{Find } \langle P^{\ast},Q^{\ast}\rangle=E^{\ast}_{\psi}[C] \text{ for which}\\
\pcind \ker(w^{\ast})=\langle P^{\ast}+[\alpha]Q^{\ast}\rangle\\
%\pcind\gamechange{$e_{C}(P,Q)=e_{C}(P^{\ast},Q^{\ast})^{B}$}\\
\text{Set } S^{\ast}:=(P^{\ast},Q^{\ast})\\
\pi^{\ast}_{\psi'}\leftarrow \mathcal{S}((E_w, E_1),1)\\
\tilde{\mathsf{\Sigma}}:=(E_1, \pi^{\ast}_{\psi'}, E^{\ast}_{\psi},S^{\ast},\mathcal{R}^{\ast}_{\tilde{\sigma}})\\
\mathsf{\Sigma}^{\ast}\leftarrow\adv^{\mathcal{H},\mathcal{O}_{S},\mathcal{O}_{pS}}(\tilde{\mathsf{\Sigma}})\\
\gamechange{$\pcreturn (m^{\ast},\mathsf{\Sigma}^{\ast})$}
    }
    \end{pcvstack}

    \begin{pcvstack}[space=.3em]
    
    \procedure[linenumbering]{$\mathcal{H} (x)$}{%
    \pcif H[x]=\perp\\
    \pcind\gamechange{$H[x]\leftarrow\mathcal{H}^{\mathsf{SQIsignHD}}(x)$}\\
    \pcreturn H[x]
    }
    \procedure[linenumbering]{$\mathcal{O}_{pS} (m,I_w)$}{%
    \text{Parse } I_w \text{ as } (E_w,w(\frak{B}),\pi_w)\\
    w := \mathcal{E}(E_w,w(\frak{B}), \pi_w , H)\\
    \pcif (w, I_w)\not\in \mathsf{R_{\mathfrak{A}}}\\
    \pcind \pcabort\\
\gamechange{$\mathsf{\Sigma}\leftarrow\mathsf{Sig^{SQIsignHD}}(m)$}\\
\text{Parse }\mathsf{\Sigma} \text{ as } (E_{1},\mathcal{R}_{\sigma})\\
\text{Extract } (E^{\ast}_{\psi},\mathcal{R}^{\ast}_{\tilde{\sigma}}) \text{ by } \mathcal{A}_{\mathsf{SIDH}} \text{ s.t. }\\
\pcind \sigma = \tilde{\sigma}\circ \hat{w}^{\ast} \text{ with } \deg(\hat{w}^{\ast})=C\\
\text{Extract }\alpha \text{ from }\\
\pcind \ker(w)=\langle P+[\alpha]Q\rangle\\
\text{Find } \langle P^{\ast},Q^{\ast}\rangle=E^{\ast}_{\psi}[C] \text{ for which}\\
\pcind \ker(w^{\ast})=\langle P^{\ast}+[\alpha]Q^{\ast}\rangle\\
%\pcind\gamechange{$e_{C}(P,Q)=e_{C}(P^{\ast},Q^{\ast})^{B}$}\\
\text{Set } S^{\ast}:=(P^{\ast},Q^{\ast})\\
\pi^{\ast}_{\psi'}\leftarrow \mathcal{S}((E_w, E_1),1)\\
\mathcal{Q}:=\mathcal{Q}\cup\{m\}\\
\pcreturn \tilde{\mathsf{\Sigma}}:=(E_1, \pi^{\ast}_{\psi'}, E^{\ast}_{\psi},S^{\ast},\mathcal{R}^{\ast}_{\tilde{\sigma}})
    }
    \procedure[linenumbering]{$\mathcal{O}_S \big(m)$}{%
    \gamechange{$\mathsf{\Sigma}\leftarrow\mathsf{Sig^{SQIsignHD}}(m)$} \\
    \mathcal{Q}:=\mathcal{Q}\cup\{m\} \\
    \pcreturn \mathsf{\Sigma}
    }
    \end{pcvstack}
\end{pchstack}

\noindent\textbf{Simulation of Oracle Queries:}\\

\noindent\textbf{Signing queries.} 
If the adversary $\mathcal{A}$ queries the signing oracle $\mathcal{O}_{S}$ with input $m$, the simulator $\mathcal{S}$ forwards $m$ to its oracle $\mathsf{Sig}^{\mathsf{SQIsignHD}}$ and then sends the response back to $\mathcal{A}$.

\noindent\textbf{Random Oracle queries.} If  $\mathcal{A}$ queries the oracle $\mathcal{H}$ with input $x$, and if $H[x] = \perp$, the simulator $\mathcal{S}$ queries $\mathcal{H}^{\mathsf{SQIsignHD}}(x)$. Otherwise, it returns $H[x]$.

\noindent\textbf{Pre-Signing queries.} When $\mathcal{A}$ submits a query $(m, I_w)$ to the pre-signing oracle $\mathcal{O}_{pS}$,
\begin{enumerate}
    \item  The simulator uses the extractability property of $\mathsf{NIZK}$ to extract the witness isogeny $w$. It then sends the message $m$ to the oracle $\mathsf{Sig}^{\mathsf{SQIsignHD}}$ and parses the resulting signature $\mathsf{\Sigma}$ as $(E_1, \mathcal{R}_{\sigma})$.
    \item The simulator $\mathcal{S}$ constructs the pre-signature isogeny representation $\mathcal{R}_{\tilde{\sigma}}$ and the torsion basis $S^{\ast} = (P^{\ast}, Q^{\ast})$ by decomposing $\sigma$ into $\tilde{\sigma} \circ \hat{w}^{\ast}$ using the algorithm $\mathcal{A}_{\mathsf{SIDH}}$, and by extracting the value $\alpha$ from the witness $w: E_0 \to E_0 / \langle P + [\alpha]Q \rangle$ obtained from the online extractor property.
    \item The simulator $\mathcal{S}$ generates a zero-knowledge proof $\pi^{\ast}_{\psi'}$ for the statement $E_1$. It then outputs the pre-signature $\tilde{\mathsf{\Sigma}} = (E_1, \pi^{\ast}_{\psi'}, E^{\ast}_{\psi}, S^{\ast}, \mathcal{R}^{\ast}_{\tilde{\sigma}})$.
\end{enumerate}

\noindent\textbf{Challenge phase:}
\begin{enumerate}
\item When $\mathcal{A}$ outputs the challenge message $(m^{\ast}, I_{w})$, the simulator $\mathcal{S}$ extracts $w$ using the extractability property of \textsf{NIZK}, sends $m^{\ast}$ to the oracle $\mathsf{Sig}^{\mathsf{SQIsignHD}}$, and parses the generated signature as $\mathsf{\Sigma} = (E_1, \mathcal{R}_{\sigma})$.
\item $\mathcal{S}$ constructs the required pre-signature $\tilde{\mathsf{\Sigma}}$ in the same way it does for $\mathcal{O}_{pS}$ queries.
\item When $\mathcal{A}$ produces a forgery $\mathsf{\Sigma}^\ast$, the simulator returns $(m^\ast, \mathsf{\Sigma}^\ast)$ as its own forgery.
\end{enumerate}

\noindent The key distinction between the simulation and $\mathsf{Game_4}$ is syntactical. Instead of generating the secret/public keys and executing the algorithms $\mathsf{Sig}$ and $\mathcal{H}$, the simulator $\mathcal{S}$ relies on its oracles $\mathsf{Sig}^{\mathsf{SQIsignHD}}$ and $\mathcal{H}^{\mathsf{SQIsignHD}}$. It remains to show that the forgery produced by $\mathcal{A}$ can be used by the simulator to win the $\mathsf{StrongSigForge}$ game.

\begin{claim}
    $(m^\ast,\mathsf{\Sigma}^{\ast})$ constitutes a valid forgery in the $\mathsf{StrongSigForge}$ game.
    \begin{proof}
        It suffices to demonstrate that the pair $(m^\ast, \mathsf{\Sigma}^\ast)$ has not been previously output by the oracle $\mathsf{Sig}^{\mathsf{SQIsignHD}}$. Note that neither $\mathcal{O}_{pS}$ nor $\mathcal{O}_{S}$ has received a query from adversary $\mathcal{A}$ on the challenge message $m^\ast$. Consequently, $\mathsf{Sig}^{\mathsf{SQIsignHD}}$ is queried on $m^\ast$ only during the challenge phase. If adversary $\mathcal{A}$ produces a forgery $\mathsf{\Sigma}^\ast$ that matches the signature $\mathsf{\Sigma}$ generated by $\mathsf{Sig}^{\mathsf{SQIsignHD}}$ during the challenge phase, the extracted $w$ would satisfy the relation with the corresponding statement $I_w$. Therefore, $\mathsf{Sig}^{\mathsf{SQIsignHD}}$ has never previously output $\mathsf{\Sigma}^\ast$ on query $m^\ast$. Thus, $(m^\ast, \mathsf{\Sigma}^\ast)$ constitutes a valid forgery in the $\mathsf{StrongSigForge}$ game.
    \end{proof}
\end{claim}

\noindent From $\mathsf{Game_0}$ to $\mathsf{Game_4}$, we have 
$$ \text{Pr}[\mathsf{Game_0} = 1] \leq \text{Pr}[\mathsf{Game_4} = 1] + \mathsf{negl}(\lambda). $$
Since $\mathcal{S}$ perfectly simulates $\mathsf{Game_4}$, it follows that we obtain:
\begin{align*}
    \mathsf{Adv}_{\mathcal{A}}^{\mathsf{aWitExt}} &= \text{Pr}[\mathsf{Game_0} = 1]\\
    &\leq \text{Pr}[\mathsf{Game_4} = 1] + \mathsf{negl}(\lambda)\\
    &\leq \mathsf{Adv}_{\mathcal{S}}^{\mathsf{StrongSigForge}} + \mathsf{negl}(\lambda).
\end{align*}

\noindent Since SQIsignHD is secure in the random oracle model with $\mathcal{H}^{\mathsf{SQIsignHD}}$ modeled as a random oracle, it follows that the adaptor signature scheme $\Xi_\mathsf{R_{\mathfrak{A}},\Sigma_{SQIsignHD}}$ achieves witness extractability in the random oracle model. This completes the proof of Lemma \ref{lemma:securityLemma4}.
    \end{proof}
\end{lemma}

\begin{theorem}\label{thm:security}
    If the SQIsignHD signature scheme, $\mathsf{\Sigma_{SQIsignHD}}$, is $\mathsf{SUF}$-$\mathsf{CMA}$-secure, $\mathsf{R_{\mathfrak{A}}}$ is a hard relation,  Problem \ref{problem:SSIP-A} and Problem \ref{problem:SSIP-B} are computationally hard, then the $\mathsf{SQIAsignHD}$ adaptor signature scheme, $\Xi_\mathsf{R_{\mathfrak{A}},\Sigma_{SQIsignHD}}$, introduced in Algorithm \ref{alg:SQIAsignHD}, is secure in the random oracle model.
    \begin{proof}
        By Lemmas \ref{lemma:securityLemma1}, \ref{lemma:securityLemma2}, \ref{lemma:securityLemma3}, and \ref{lemma:securityLemma4}, we have demonstrated that the adaptor signature scheme $\Xi_\mathsf{R_{\mathfrak{A}}, \Sigma_{SQIsignHD}}$ satisfies the properties of pre-signature correctness, pre-signature adaptability, $\mathsf{aEUF\text{-}CMA}$ security, and witness extractability. The verification of these properties completes the proof of Theorem \ref{thm:security}.
    \end{proof}
    \label{securitytheorem}
\end{theorem}

\section*{Conclusion}
Adaptor signatures, an extension of standard digital signatures, are a vital cryptographic primitive for blockchain applications, helping to reduce costs, enhance fungibility, and support off-chain payments within payment-channel networks and hubs. In the present work, we have introduced $\mathsf{SQIAsignHD}$, a new adaptor signature construction with quantum-resistant security based on isogenies of supersingular elliptic curves. Thereby, it provides security and privacy concepts relevant to off-chain applications. In $\mathsf{SQIAsignHD}$, we use SQIsignHD as the underlying signature scheme and make use of the idea of artificial orientation, on the supersingular isogeny Diffie-Hellman key exchange protocol (SIDH), to apply the hard relation. We also exploit the SIDH attacks as a generic algorithm in recovering the secret witness isogeny in the extraction phase of our scheme. The signature in $\mathsf{SQIAsignHD}$ is approximately $1.26$ KB in size for $\lambda=128$ security level. In contrast to the only isogeny-based adaptor signature construction, IAS, which operates on a maximum of the CSIDH-512 parameters, our scheme scales well to high-security levels. Thus, compared to IAS, $\mathsf{SQIAsignHD}$ significantly improves the security level and signature size. Providing a concrete and optimized implementation of $\mathsf{SQIAsignHD}$ is left for future work.

%\section*{Statements and Declarations}
%The authors would like to declare that there are no known conflicts of interest or personal relationships linked with this work and there has been no financial support for this work that could have influenced its outcome.


%\bibliographystyle{plain}
%\bibliography{Bibliography.bib}
\end{document}